\documentclass[a4paper,fleqn,usenatbib]{mnras}

\usepackage{newtxtext,newtxmath}

\usepackage{longtable}
\usepackage[T1]{fontenc}
\usepackage{ae,aecompl}

\usepackage{graphicx}	
\usepackage{amsmath}	
\usepackage{amssymb}	
\usepackage{color}
\usepackage[english]{babel}
\usepackage{graphicx}
\usepackage{hyperref}
\usepackage{listings}
\usepackage{lscape}
\usepackage{natbib}
\usepackage{url}
\usepackage{breakurl}
\usepackage{xspace}
\bibpunct{(}{)}{;}{a}{}{,}

\newcounter{Rco}

\newcommand{\Ionst}[1]{\setcounter{Rco}{#1}\Roman{Rco}}
\newcommand{\Ion}[2]{\mbox{#1\,{\scriptsize\Ionst{#2}}}}

\newcommand{\logg}{\mbox{$\log g$}\xspace}
\newcommand{\loggw}[1]{\mbox{$\log g\hspace{-0.5mm} =\hspace{-0.5mm}  #1$}}

\newcommand{\ab}[1]{\mbox{Fig.\,\ref{#1}}}
\newcommand{\sA}[1]{\mbox{(Fig.\,\ref{#1})}}

\newcommand{\ta}[1]{\mbox{Table\,\ref{#1}}}
\newcommand{\sT}[1]{\mbox{(Table\,\ref{#1})}}
\newcommand{\Teff}{\mbox{$T_\mathrm{eff}$}\xspace}
\newcommand{\Teffw}[1]{\mbox{$\Teff\hspace{-0.5mm} =\hspace{-0.5mm} #1 \,\mathrm{K}$}}
\newcommand{\ebv}{\mbox{$E_{B-V}$}}
\newcommand{\ebvw}[1]{\mbox{$\ebv\hspace{-0.5mm} =\hspace{-0.5mm} #1$}}

\newcommand{\nh}{\mbox{$N_\ion{H}{i}$}}
\newcommand{\nhw}[1]{\mbox{$\nh\hspace{-0.5mm} =\hspace{-0.5mm} #1\, \mathrm{cm}^{-2}$}}

\newcommand{\Msol}{$M_\odot$}
\newcommand{\bd}{BD$-22\degr 3467$\xspace}
\newcommand{\gb}{G191$-$B2B\xspace}
\newcommand{\re}{RE\,0503$-$289\xspace}
\newcommand{\wdn}{WD\,0111$+$002\xspace}
\newcommand{\pgs}{PG\,1707$+$427\xspace}
\newcommand{\pgn}{PG\,0109$+$111\xspace}
\title[Discovery of trans-iron elements in \bd]
      {First discovery of trans-iron elements \\ in a DAO-type white dwarf (\bd)}

\author[L\@. L\"obling et al.]{
L\@. L\"obling$^{1}$\thanks{E-mail: loebling@astro.uni-tuebingen.de},
M.\,A\@. Maney$^{1,2}$,
T\@. Rauch$^{1}$,
P\@. Quinet$^{3,4}$, \newauthor
~S\@. Gamrath$^{3}$,
J\@. W\@. Kruk$^{5}$,
and
K\@. Werner$^{1}$
\\
$^{1}$Institute for Astronomy and Astrophysics,
           Kepler Center for Astro and Particle Physics,
           Eberhard Karls University,\\ 
           Sand 1,
           72076 T\"ubingen, 
           Germany\\
$^{2}$ Department of Astronomy \& Astrophysics, Eberly College of Science, The Pennsylvania State University, 525 Davey Lab,\\
University Park, PA 16802, USA\\
$^{3}$Physique Atomique et Astrophysique, Universit\'e de Mons -- UMONS, 7000 Mons, Belgium\\
$^{4}$IPNAS, Universit\'e de Li\`ege, Sart Tilman, 4000 Li\`ege, Belgium\\
$^{5}$NASA Goddard Space Flight Center, Greenbelt, MD\,20771, USA}
\date{Accepted 2019 November 18. Received 2019 November 11; in original form 2019 September 16}

\pubyear{2019}
\begin{document}
\label{firstpage}
\pagerange{\pageref{firstpage}--\pageref{lastpage}}
\maketitle

\begin{abstract}
We have identified 484 lines of the trans-iron elements (TIEs) Zn, Ga,
Ge, Se, Br, Kr, Sr, Zr, Mo, In, Te, I, Xe, and Ba, for the first time
in the ultraviolet spectrum of a DAO-type WD, namely \bd, surrounded
by the ionized nebula Abell\,35.  Our TIE abundance
determination shows extremely high overabundances of up to five dex --
a similar effect is already known from hot, H-deficient (DO-type)
white dwarfs. In contrast to these where a pulse-driven convection
zone has enriched the photosphere with TIEs during a final thermal
pulse and radiative  levitation has established the extreme TIE
overabundances, {here the extreme TIE overabundances 
are exclusively driven by radiative levitation on the initial stellar metallicity. 
The very low mass ({$0.533^{+0.040}_{-0.025}\,M_\odot$}) of \bd implies that a 
third dredge-up with enrichment of s-process elements in the photosphere did not 
occur in the AGB precursor.}
\end{abstract}




\section{Introduction}
\label{sect:intro}
Trans-iron elements (TIEs) are synthesized during the asymptotic giant
branch (AGB) phase of a star by the slow neutron-capture (s-)process. 
Depending on the initial stellar mass, its yields vary
strongly \citep{karakasetal2016}. To become detectable, TIEs have to
be transported from the helium-rich intershell region to the stellar
surface. This happens, if the star experiences a third dredge-up {\citep[TDU, c.f.,][]{2000A&A...360..952H}}.
A scenario in which the envelope becomes mixed with the intershell
region  is known as the late helium-shell flash. Such a late thermal
pulse (LTP) was predicted, e.g., by \citet{ibenetal1983}. When it
occurs after the star's descent from the AGB at already declining
luminosity, i.e., close to the end of nuclear burning, the H-burning
shell is ``off'' and a pulse-driven convection zone (PDCZ) establishes
between the He-burning shell and the photosphere.  The remaining H is
mixed into the stellar interior, becomes diluted or even burned,
making the star H-deficient {\citep[c.f.,][]{1977PASJ...29..331F, 1979A&A....79..108S, ibenetal1983, 1995A&A...299..755B}}.  Thus, it was -- although surprising --
well understandable, that lines of ten TIEs were identified
\citep{werneretal2012} in the ultraviolet (UV) spectrum of the
DO-type white dwarf (WD) \re  \citep[effective temperature
  \Teffw{70\,000 \pm 2000}, surface gravity
  $\log\,(g\,/\,\mathrm{cm\,s^{-2}}) = 7.5 \pm
  0.1$,][]{rauchetal2016kr},  which became an archetype for TIE search
in WDs. Presently, 18 of these species are identified in \re (Rauch et
al\@. submitted).  \citet{chayer2005} first succeeded in the detection
of TIEs in DO WDs, namely six species in two other objects
(HD\,149499\,B, HZ\,21).

In a subsequent investigation, TIE line identification was
successfully performed in three related H-deficient objects
\citep[two DO-type WDs and one PG\,1159-type WD, namely \wdn, \pgn,
  and \pgs,][]{hoyeretal2018}.  The commonality of these stars is that
they are located close to the so-called PG\,1159 wind limit
\citep{unglaubbues2000} that approximately separates  the regions of
PG\,1159-type stars and DO-type WDs in the Hertzsprung-Russell diagram
(HRD).  Here, the stellar wind is already weak enough and diffusion
can establish strong TIE overabundances of up to five dex in the
photosphere \citep{rauchetal2016mo}.

The search for TIE lines has not been restricted to {He-rich} WDs. \citet{vennes2005} discovered the first TIE in WDs at all, namely
Ge in three DA WDs. One of them is \gb, an object  that is employed as
spectrophotometric flux standard for the Hubble Space Telescope
\citep[e.g.,][]{bohlin2007,rauchetal2013}. Recently, the TIEs Cu, Zn,
Ga, Ge, As, Sn, and Ba were identified \citep[Rauch et
  al\@. submitted,][]{ rauchetal2014zn, rauchetal2015ga,
  rauchetal2012ge, rauchetal2016mo, rauchetal2013, rauchetal2014ba}.
The TIE abundance pattern is similar to \re, but at a lower absolute
level probably because of the lower \Teff of \gb
\citep[\Teffw{60\,000}, ][]{rauchetal2013}.  TIE line search and
abundance analyses are also successfully  performed in the field of
He-rich, hot subdwarf stars.
{The first one was {LS\,IV$-14\degr 116$\xspace}, for which extreme overabundances of Fe, Sr, Y, and Zr were detected \citep{2011MNRAS.412..363N}. The most recent members of the group of ``heavy metal'' subdwarfs are }
HZ\,44, HD\,127493, and Feige\,46
\citep[e.g., ][]{dorschetal2019,latouretal2019}, with TIE enrichment
patterns similar to those in Fig.\,\ref{fig:x}.

{Abell\,35 was discovered by \cite{1955PASP...67..258A} and characterized as homogeneous disk planetary nebula \cite[PN, ][]{1966ApJ...144..259A}. \cite{1981ApJ...244..903J} classified the visible nucleus as G8 III$-$IV. Later, \cite{1988ESASP.281b.177G} classified the hot, ionizing central star as DAO-type white dwarf (WD). Shortwards of 2800\,{\AA}, the WD dominates the flux, whereas the cool companion outshines it in the optical. \cite{2002ApJ...580..434H} analyzed the binary and found \Teffw{80\,000} and $\log\,(\,g\,/\,\mathrm{cm}/\mathrm{s}^2\,) = 7.7$  for the hot and \Teffw{5\,000} and \loggw{3.5} for the cool star. \cite{ziegleretal2012} corrected the surface gravity of the WD to \loggw{7.2} and \cite{frewetal2010} classified the nebula as ``bow shock nebula in a photoionized Strömgren sphere''.}
Recently, a close re-inspection of the UV spectrum of the exciting
star of the ionized nebula Abell\,35, \bd
\citep[WD\,1250-226,][]{mccooksion1999}, led us to
the identification of TIE absorption lines. In this work, a systematic
TIE line search was  performed in order to constrain abundances
analogously to  \citet[for \re]{hoyeretal2017}. It is based on the \bd
model of \citet[][atmospheric parameters given in
  Table\,\ref{tab:finab}]{ziegleretal2012} {that was calculated using the non-local thermodynamic
equilibrium (NLTE) model-atmosphere code of the T\"ubingen NLTE Model
Atmosphere Package
\citep[TMAP\footnote{\url{http://astro.uni-tuebingen.de/~TMAP}},][]{werneretal2003,tmap2012}}.  In
Sects.\,\ref{sect:observations} and \ref{sect:models}, we briefly
describe the available observations and the model atmospheres used for
the spectral analysis, respectively.  In
Sect.\,\ref{sect:linesabundances}, the process of line identification
and subsequent abundance measurement is explained. Lastly, we
summarize our results and conclude in Sect.\,\ref{sect:results}.

\section{Observations}
\label{sect:observations}

Our analysis is based on high-resolution Far Ultraviolet Spectroscopic
Explorer (FUSE) and Space Telescope  Imaging Spectrograph (HST/STIS)
observations. These were obtained from the
MAST\footnote{http://archive.stsci.edu} archive. {The FUSE spectrum taken with the LWRS aperture has a resolving power of $R = \lambda\,/\,\Delta\lambda \approx 20\,000$. Four STIS observations with grating E140M and $R = 45\,800$ are available. The
observation log is shown in \ta{tab:obslog}.}
{We convolved the synthetic spectra} with Gaussians to model the respective instrument's resolution. The signal-to-noise
ratio of the STIS observations was improved by co-adding the observations.
{The combined spectra are the same as used by \citet{ziegleretal2012}. No optical observation of the DAO WD are available, since the G-star companion dominates this spectral range.}

\section{Model atmospheres and atomic data}
\label{sect:models}

The analysis was carried out using {TMAP}. 
This code assumes plane-parallel geometry and calculates chemically
homogeneous NLTE atmospheres in radiative and hydrostatic equilibrium. 

For the TIEs Cu, Zn, Ga, Ge, Se, Br, Kr, Sr, Zr, Mo, In, Te, I, Xe,
and Ba, we used the recently calculated data that is available via the
T\"ubingen Oscillator Strengths Service (TOSS). For the elements with
$Z \ge 20$, it is necessary to create model atoms using a statistical
approach that calculates super levels and super lines
\citep{rauchdeetjen2003} to take their complex atomic structure into
account for the calculation. The statistics of all elements
considered in our model-atmosphere calculations are summarized in
Table\,\ref{tab:stat}. 

We constructed a new classical model ion for \Ion{Ba}{8} from the
level and line data of \citet{churilov2001} available via the
National Standards and Technology Institute (NIST) Atomic Spectra
Database
\citep[ASD\footnote{\url{https://physics.nist.gov/PhysRefData/ASD}},
][]{NIST_ASD}, which was incorporated into TMAD.

For all considered elements with an atomic number $Z \le 28$, we used
the same model atoms like \citet{ziegleretal2012}. For $Z <
20$ these were obtained from the T\"ubingen Model Atom
Database \citep[TMAD,][]{rauchdeetjen2003} that was constructed as
part of the German Astrophysical Virtual Observatory (GAVO). For the
iron-group elements (IGEs, atomic number $20 \le Z \le 28$), Kurucz's
line lists\footnote{\url{http://kurucz.harvard.edu/atoms.html}}
\citep{kurucz2009,kurucz2011,kurucz2017} were utilized. 

For this analysis, we adopted the photospheric parameters of
\citet{ziegleretal2012} and used their final model (\Teffw{80\,000},
\loggw{7.2}, see \ta{tab:finab} for the element abundances) to start
our TIE analysis.  To identify lines and determine abundances of the
15 TIEs {(Cu, Zn, Ga, Ge, Se, Br, Kr, Sr, Zr, Mo,
In, Te, I, Xe, and Ba)}, we performed
line-formation calculations {by adding each of them individually to the start model from \citet{ziegleretal2012}}, while temperature and density
structure of the atmosphere were kept fixed.  To verify our method, a
final model including all TIEs with their previously determined
abundances was then  calculated with temperature and density
corrections. The deviations in the abundances were marginal.  

{The observed spectra are affected by reddening due to interstellar material within the line of sight.
By comparing the slope of the flux calibrated observations as well as GALEX, HIPPARCOS, and 2MASS magnitudes \citep{binachietal2011,perryman1997,cutrietal2003} with the synthetic spectra of the central star, \citet{ziegleretal2012} found a color excess \ebvw{0.02 \pm 0.02}. This value} was used to apply interstellar reddening
following the law of \citet[][with the standard $R_\mathrm{V} =
  3.1$]{fitzpatrick1999} to the model spectra to reproduce the
observation. Absorption due to neutral interstellar
hydrogen, assuming a column density of \nhw{5.0 \times
10^{20}} \citep{ziegleretal2012}, was applied to the synthetic
spectra. Furthermore, we applied the interstellar line-absorption
model of \citet{ziegleretal2012} that was calculated using
the program OWENS \citep{hebrardetal2002,hebrardetal2003} to
unambiguously identify lines of stellar and interstellar origin.

\section{Line identification and abundance determination}
\label{sect:linesabundances}

We calculated synthetic spectra from our line-formation models with
each of the 15 elements added individually to the best model of
\citet{ziegleretal2012}. 
{The spectrum is crowded with a multitude of blended metal lines which hampers their unambiguous identification. 
To clearly see the contribution of the individual TIE elements, we divided the synthetic spectrum including this species by another model spectrum without it.
The individual abundances were varied by small steps of 0.2\,dex or smaller to derive the final values from evaluation of line-profile fits by eye.
To estimate the influence of the uncertainty in \Teff of $\pm
10\,000$\,K and in \logg of $\pm 0.3$ for the error propagation, we
redid the abundance determination for models with \Teffw{90\,000} and
\loggw{6.9} as well as for \Teffw{70\,000} and \loggw{7.5}. The
abundances are affected by typical errors below 0.3\,dex.}

For Cu, a line
identification was not possible with  appreciable certainty. Instead,
upper limits were determined by reducing the abundance until the
strongest computed lines become undetectable. An equivalent width of
$W_\lambda = 5$\,m{\AA} was set as a detection
limit. \ta{tab:Culines}\,$-$\,\ta{tab:Balines} list all lines of TIEs,
that appear with an equivalent width above the threshold in the model
spectrum. These tables include also those lines that
could not be identified in the spectrum of \bd due to, e.g., blending
with other photospheric or interstellar lines to make them a useful
tool for the identification of TIE lines in the spectra of other
DAO-type WDs. 
{The abundances are given in \ta{tab:finab} and are illustrated in \ab{fig:x}}.
The complete FUSE and STIS observations compared to our
best model are shown
online\footnote{\url{http://astro.uni-tuebingen.de/~TVIS/objects/Abell35}}
within the T\"ubingen VISualization tool (TVIS). The ionization
fractions as well as the temperature structure and electron density in
the final atmosphere model are shown in \ab{fig:ionfrac}.

The number of identified lines per TIE ion is shown in
\ta{tab:tielines}. The observation is well reproduced by our final
model with the abundances shown in \ta{tab:finab} as it is illustrated
in \ab{fig:zn} to \ref{fig:ba} for prominent lines of each of the
TIEs.

\begin{table}\centering
\caption{Numbers of identified lines in the ionization stages \textsc{iv-viii} of TIEs in the UV spectrum of \bd.}
\label{tab:tielines}
\begin{tabular}{rrccccc}
\hline\hline
Element & $Z$ & \textsc{iv} & \textsc{v} & \textsc{vi} & \textsc{vii} & \textsc{viii} \\
\hline
Zn & 30 & $2$ &$141$& $  $ & $ $ & $ $ \\
Ga & 31 & $2$ &$71$ & $52$ & $ $ & $ $ \\
Ge & 32 & $2$ &$32$ & $57$ & $ $ & $ $ \\
Se & 34 & $ $ &$14$ & $  $ & $ $ & $ $ \\
Br & 35 & $ $ &$1 $ & $7 $ & $ $ & $ $ \\
Kr & 36 & $ $ &$  $ & $4 $ & $ $ & $ $ \\
Sr & 38 & $ $ &$17$ & $  $ & $ $ & $ $ \\
Zr & 40 & $ $ &$ 1$ & $28$ & $3$ & $ $ \\
Mo & 42 & $ $ &$  $ & $2 $ & $ $ & $ $ \\
In & 49 & $ $ &$28$ & $  $ & $ $ & $ $ \\
Te & 52 & $ $ &$  $ & $3 $ & $ $ & $ $ \\
I  & 53 & $ $ &$  $ & $4 $ & $ $ & $ $ \\
Xe & 54 & $ $ &$  $ & $  $ & $4$ & $ $ \\
Ba & 56 & $ $ &$  $ & $  $ & $3$ & $3$ \\
\hline
\end{tabular}
\end{table}

\begin{figure*}
  \resizebox{\hsize}{!}{\includegraphics{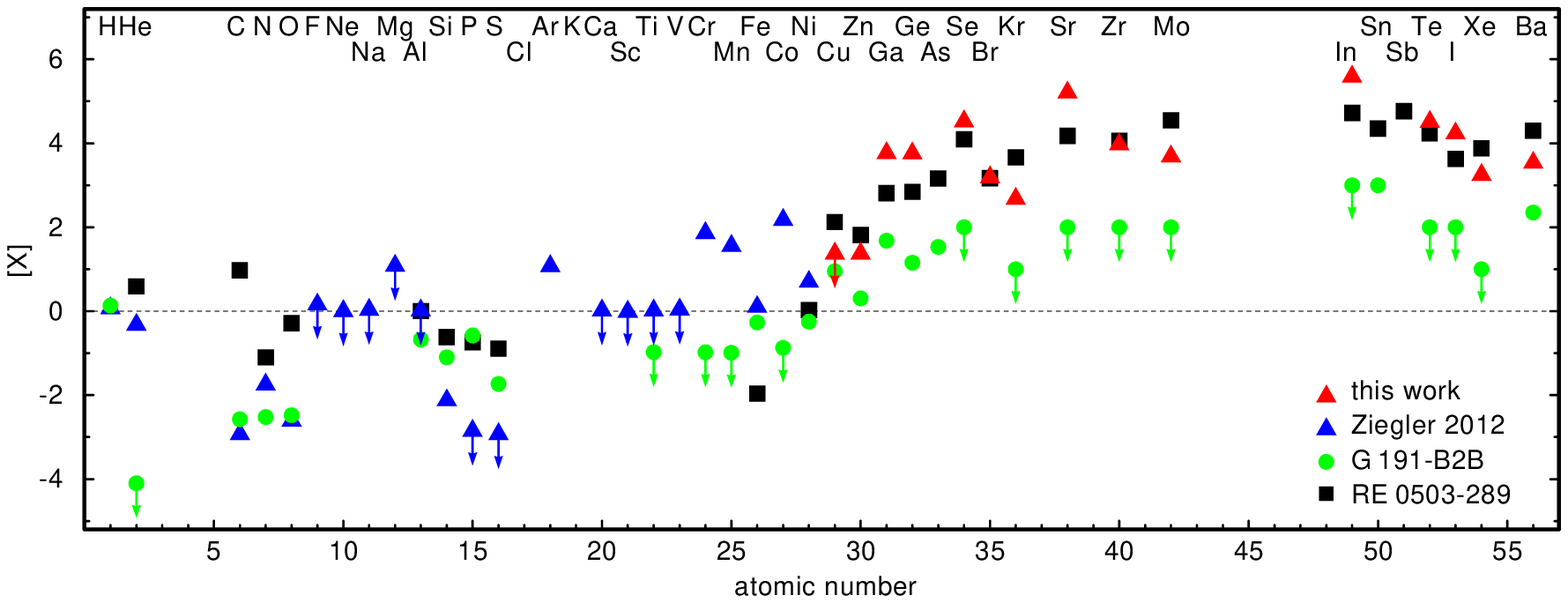}}
   \caption{Photospheric abundance ratios [X] = log\,(mass
     fraction/solar mass fraction) of \bd determined from detailed
     line profile fits. Solar values are taken from \citet{asplundetal2009,scottetal2015a,scottetal2015b,grevesseetal2015}. 
     Upper limits are indicated with arrows. The
     black, solid line indicates solar abundances. 
     Blue triangles represent the abundances determined by \citet{ziegleretal2012}, 
     red triangles show the TIE abundances \sT{tab:finab}. 
     For comparison, the abundances determined for 
     \gb \citep[green circles]{rauchetal2013} and
     \re are shown \citep[black squares]{hoyeretal2017}.}
   \label{fig:x}
\end{figure*}

\begin{figure}
  \resizebox{\hsize}{!}{\includegraphics{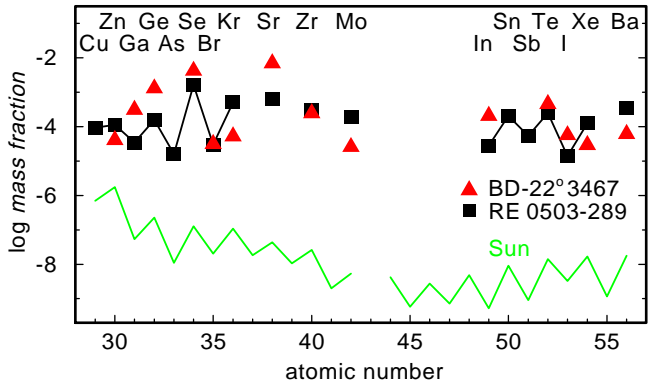}}
   \caption{Photospheric TIE abundances in \bd (red triangles) compared to \re \citep[black squares][Rauch et al\@. 2019 submitted]{hoyeretal2017}. 
    Solar values are shown for comparison.}
   \label{fig:massfractionre}
\end{figure}

\section{Results and conclusions}
\label{sect:results}

To identify TIE lines, the UV spectrum of \bd was closely inspected
which led to the discovery of  Zn, Ga, Ge, Se, Br, Kr, Sr, Zr, Mo, In,
Te, I, Xe, and Ba (Table\,\ref{tab:tielines}).  In total, 484 TIE
lines were discovered.

Our spectral analysis has shown that the enrichment of TIEs in  \bd
(\Teffw{80\,000 \pm 10\,000}, \loggw{7.2 \pm 0.3}) and  \re
(\Teffw{70\,000 \pm  2000}, \loggw{7.5 \pm 0.1}) is comparably high
($\approx 1.5 - 5\,\mathrm{dex}$, Fig.\,\ref{fig:x}).  The origin of
the high enrichment of TIEs is diffusion, i.e., efficient radiative
levitation. This was shown already for \re  by detailed diffusion
calculations \citep{rauchetal2016mo}. While it was possible to
determine abundances  for several TIEs with  consecutive atomic number
in \re and find that the odd-even shape of the solar abundance pattern
seems to be reflected also by the  enriched TIEs
(\ab{fig:massfractionre}, cf\@., Rauch et al\@. 2019, submitted),
there is not enough information to confirm  this finding based on the
results for \bd. 

The evolutionary difference between \bd and \re is that the latter
most likely experienced an LTP in which it became
hydrogen-deficient.
{As a result, a pulse-driven convection zone, established during the flash, enriched the TIEs in the atmosphere. Their abundances were later on amplified to the observed values by radiative levitation. In contrast, the high abundances of TIEs in \bd are possibly the result of radiative levitation on the initial stellar metallicity without previous enrichment by s-processed matter. This is because of the very low mass of \bd, (Fig.\,\ref{fig:x}) which corresponds to an initial mass of below $1.0\,M_\odot$ \citep{cummingsetal2018}, implying that no TDU occurred on the AGB \citep{karakasetal2016}. From the position in the \Teff-\logg diagram \sA{fig:loggteff} only, a possible evolution without an AGB phase, directly from the extended horizontal branch (EHB) to the WD cooling sequence, cannot be excluded. Conversely, it is then possible that the high amount of TIEs in \re is solely due to diffusion, independent of the occurrence of a previous LTP. This is an interesting conclusion, because a large fraction of DOs is not initiated by an LTP but by a merger event with the so-called O(He) stars as merger products and DO precursors \citep{reindl2014}. DOs from this evolutionary channel should therefore also go through a phase with an extreme TIE enrichment by diffusion.}
%

We have to keep in mind that
diffusion-established abundance patterns do not contain anymore
information about the previous stellar evolution, i.e., wherever the
TIEs (or other elements) stem from, the exhibited surface abundance
may be the same.  Investigations on yields of the AGB s-process
nucleosynthesis elements have to be performed before diffusion
dominates the stellar evolution. This is the phase of just declining
luminosity when the strength of the stellar wind decreases {but is 
still high enough to maintain the original abundance ratios produced by the s-process. Spectral analyses of stars in that evolutionary phase might help to directly constrain AGB nucleosynthesis.}

{The formation of Abell\,35-like
central stars of planetary nebulae (CSPNe), i.e., binary CSPNe
with a rapidly rotating late-type (sub)giant and an extremely hot
companion \citep{bondetal1993}, is discussed controversially in the literature.}
\citet{theveninetal1997} found the
companion of \bd to be enriched in Ba indicating that this still
unevolved star experienced mass transfer from a (post-) AGB
star. However, this formation channel is debated since Abell\,35 was
found to only mimic a PN \citep{frewetal2010} and the mass of the
ionizing star was considered to be too low for a post-AGB star
\citep{ziegleretal2012}. {As explained above, an evolution directly from the EHB to the WD cooling track is possible \sA{fig:loggteff}.} The peculiar ionized nebula
around \bd {is not a real PN but, nevertheless, it also cannot be excluded, that the nebula material is in some way connected to the evolution of the central star.} Assuming
that it ejected a PN as an AGB star, the original PN might have
already dispersed. The star has a high proper motion
\citep[$\mu_\alpha = -54.566 \pm 0.226$\,mas/yr and  $\mu_\delta =
  -10.097 \pm 0.187$\,mas/yr,][]{Gaia2018} and, thus, might now be
passing through the edges of the ejected former nebula material or
another dense ISM region while ionizing the surrounding material. The
classification of Abell\,35 as a bow shock nebula in a photoionized
Str\"omgren sphere in the ambient ISM \citep{frewetal2010} does not
necessarily include a PN but also, does not rule out the post-AGB
nature of \bd. {Further detailed abundance analyses of a sample of Abell\,35-like CSPNe as well as their ambient nebulae and companion stars should give us a better handle on their evolution. The nebulae, if ejected from an AGB star, contain signatures of s-process elements \citep{madonna2017} as well as the unevolved companions, if they accreted a fraction of the ejected material. Therefore, a precise know\-ledge of the companion is mandatory because any accreted material would become diluted in this star's convective envelope.}

To better understand the late evolutionary phases of low-mass stars,
it is highly desirable to improve the determination of \Teff and \logg
with much narrower error ranges. For this purpose, the analysis of
high-resolution optical spectra, {may be helpful}. Although the cool
companion dominates this wavelength regime, broad lines of H and He of
\bd should be detectable {like demonstrated by \citet{aller2015} for the binary CS of NGC\,1514. These lines may reduce the error and, thus, allow to better constrain the stellar mass \sA{fig:loggteff}.} 

{As a remark, we would like to mention that the discrepancy found by \citet{ziegleretal2012} between the spectroscopic distance of $361^{+195}_{-137}\,$pc and the distance based on the HIPPARCOS parallax is still present in the era of Gaia with a distance of $124.84^{+2.21}_{-2.13}$\,pc \citep{baillerjonesetal2018}. \citet{ziegleretal2012} demonstrated, that the \Ion{H}{1} lines in the FUSE range are poorly reproduced with models with $\log g > 7.7$. This phenomenon of too large spectroscopic distances has already been reported in the literature for CSPNe \citep{2018A&A...609A.126S,2019A&A...625A.137S}. The argument that missing metal-line blanketing and back warming may result in too-high temperatures does not hold in our analysis, because all elements in the model calculation were considered in full NLTE computations. The objects of the mentioned studies are all located before the knee at highest temperatures in the Hertzsprung-Russell diagram. Thus, the spectroscopic mass derived from fitting of the spectral energy distribution using the parallax distance is systematically too high. In our case, the spectroscopic mass derived from the dereddened GALEX FUV flux \citep{binachietal2011} using the Gaia distance is unreasonably low. With the given \Teff, at least a $\logg > 8.0$ would be required to reach masses above $0.4$\,\Msol. In conclusion, the discrepancy remains unexplained and needs further investigation.}\\
Following the discovery of TIE lines in \bd, we have initiated an
analogous search in other DAO-type WDs.  Since the FUSE and
HST archives provide quite a number of high-quality UV spectra of such
stars, that have not been inspected in focus of TIEs, we expect to
identify TIE enrichment as a common phenomenon in many hot WDs.

\begin{figure}
  \resizebox{\hsize}{!}{\includegraphics{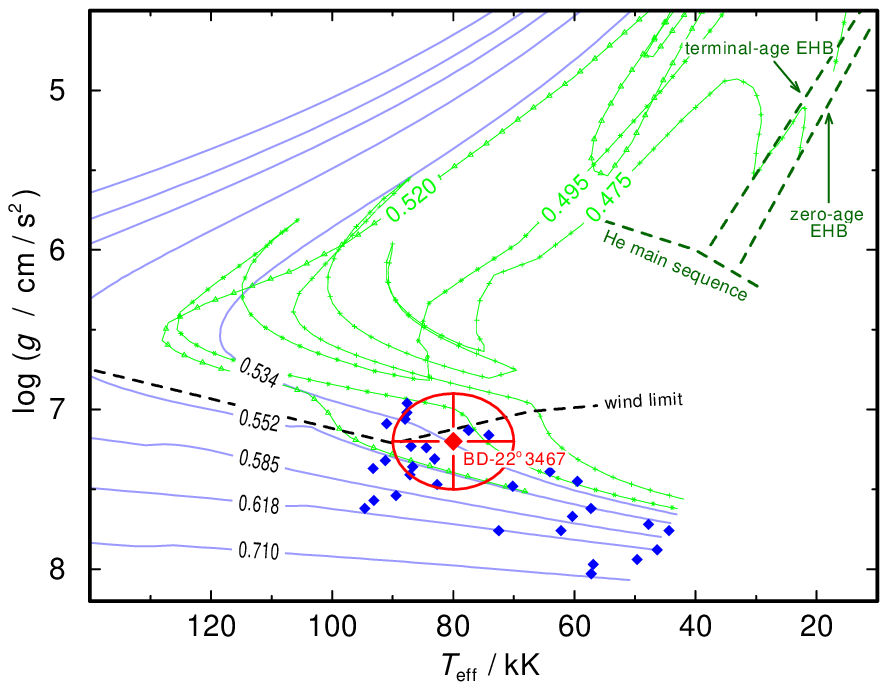}}
   \caption{Location of \bd (green) with error ellipse in the
     $T_\mathrm{eff}$-$\log g$ diagram compared to other DAO-type WDs
     \citep[][blue diamonds]{gianninasetal2010}. Included are post-EHB
     evolutionary tracks \citep[green,][$Y = 0.288 \approx
       Y_\odot$]{dormanetal1993} and post-AGB tracks for H-burners
     \citep[blue,][]{millerbertolami2016} labeled with the stellar
     mass (in $M_\odot$). The wind limit for H-rich post-AGB stars is
     indicated \citep{unglaubbues2000}.}
\label{fig:loggteff}
\end{figure}

\section*{Acknowledgments}
We thank the referee for the very useful comments that improved this paper.
LL has been supported by the German Research Foundation (DFG) under grant WE\,1312/49$-$1.
MAM had been supported by the DAAD RISE Germany program.
The GAVO project at T\"ubingen
had been supported by the Federal Ministry of Education and Research (BMBF) 
at T\"ubingen (05\,AC\,6\,VTB, 05\,AC\,11\,VTB).
Financial support from the Belgian FRS-FNRS is also acknowledged. 
PQ is research director of this organization.
Some of the data presented in this paper were obtained from the
Mikulski Archive for Space Telescopes (MAST). STScI is operated by the
Association of Universities for Research in Astronomy, Inc., under NASA
contract NAS5-26555. Support for MAST for non-HST data is provided by
the NASA Office of Space Science via grant NNX09AF08G and by other
grants and contracts. 
The 
TIRO (\url{http://astro-uni-tuebingen.de/~TIRO}),
TMAD (\url{http://astro-uni-tuebingen.de/~TMAD}),
TOSS (\url{http://astro-uni-tuebingen.de/~TOSS}), and
TVIS (\url{http://astro-uni-tuebingen.de/~TVIS}) tools and services used for this paper 
were constructed as part of the T\"ubingen project 
(https://uni-tuebingen.de/de/122430)
of the German Astrophysical Virtual Observatory
(GAVO, \url{http://www.g-vo.org}).
This research has made use 
of NASA's Astrophysics Data System and
of the SIMBAD database, 
operated at CDS, Strasbourg, France.
This work has made use of data from the European Space Agency (ESA) mission
{\it Gaia} (\url{https://www.cosmos.esa.int/gaia}), processed by the {\it Gaia}
Data Processing and Analysis Consortium (DPAC,
\url{https://www.cosmos.esa.int/web/gaia/dpac/consortium}). Funding for the DPAC
has been provided by national institutions, in particular the institutions
participating in the {\it Gaia} Multilateral Agreement.




\bibliographystyle{mnras}
\bibliography{a35_TIEs} 



\clearpage

\appendix

\section{Photospheric parameters of \bd}

\begin{table*}
\setlength{\tabcolsep}{0.80em}
\caption[]{Parameters of \bd.}
\centering
\label{tab:finab}
\begin{tabular}{r@{}lrrrrr}
\hline
\hline
\multicolumn{2}{r}{$T_\mathrm{eff}\,/\,$K}           & \multicolumn{5}{l}{$80\,000 \pm 10\,000\,^{(a)}$}  \\
\multicolumn{2}{r}{$\log\,(g$\,/\,cm/s$^2$)}        &  \multicolumn{5}{l}{$7.2 \pm 0.3\,^{(a)} $}         \\
\multicolumn{2}{r}{\ebv}                            &  \multicolumn{5}{l}{$0.02\pm 0.02\,^{(a)}$}         \\
\multicolumn{2}{r}{$\log \nh\,/\,\mathrm{cm^{-2}}$}  &  \multicolumn{5}{l}{$20.7\pm 0.1\,^{(a)}$}          \\
\multicolumn{2}{r}{$M\,/\,M_\odot$}                  & \multicolumn{2}{l}{$0.533^{+0.040}_{-0.025}\,^{(b)}$} &
                                                       \multicolumn{2}{l}{$0.505^{+0.030}_{-0.025}\,^{(c)}$}  \\
\noalign{\smallskip}
\multicolumn{2}{r}{$\log\ ( L\,/\,L_\odot )$}        & \multicolumn{2}{l}{$1.52^{+0.40}_{-0.50}\,^{(b)}$} &
                                                      \multicolumn{2}{l}{$1.69^{+0.30}_{-0.50}\,^{(c)}$}   \\      
\hline
\multicolumn{2}{r}{Element} & [X] & Mass fraction & Number fraction & $\varepsilon^{(d)}$ & [X/Fe] \\
\hline
\hbox{}\hspace{10mm}H &$\,^{(a)}$   &   $ 0.07$   &   $8.67\times 10^{-1}$   &   $9.67\times 10^{-1}$   &   $12.09$   &   $-0.00$\\
                    He&$\,^{(a)}$   &   $-0.33$   &   $1.17\times 10^{-1}$   &   $3.29\times 10^{-2}$   &   $10.62$   &   $-0.40$\\
                    C &$\,^{(a)}$   &   $-2.93$   &   $2.77\times 10^{-6}$   &   $2.59\times 10^{-7}$   &   $ 5.51$   &   $-3.00$\\
                    N &$\,^{(a)}$   &   $-1.75$   &   $1.24\times 10^{-5}$   &   $9.99\times 10^{-7}$   &   $ 6.10$   &   $-1.82$\\
                    O &$\,^{(a)}$   &   $-2.61$   &   $1.42\times 10^{-5}$   &   $9.99\times 10^{-7}$   &   $ 6.10$   &   $-2.68$\\
                    F &$\,^{(a)}$   &$\le 0.16$   &$\le5.04\times 10^{-7}$   &$\le2.98\times 10^{-8}$   &$\le 4.57$   &$\le-0.07$\\
                    Ne&$\,^{(a)}$   &$\le-0.00$   &$\le1.26\times 10^{-3}$   &$\le6.99\times 10^{-5}$   &$\le 7.94$   &$\le-0.07$\\
                    Na&$\,^{(a)}$   &$\le 0.03$   &$\le2.92\times 10^{-5}$   &$\le1.43\times 10^{-6}$   &$\le 6.25$   &$\le-0.07$\\
                    Mg&$\,^{(a)}$   &$\le 1.08$   &$\le8.31\times 10^{-3}$   &$\le3.84\times 10^{-4}$   &$\le 8.68$   &$\le 1.00$\\
                    Al&$\,^{(a)}$   &$\le 0.02$   &$\le5.56\times 10^{-5}$   &$\le2.31\times 10^{-6}$   &$\le 6.46$   &$\le-0.07$\\
                    Si&$\,^{(a)}$   &   $-2.12$   &   $5.00\times 10^{-6}$   &   $2.00\times 10^{-7}$   &   $ 5.40$   &   $-2.19$\\
                    P &$\,^{(a)}$   &$\le-2.85$   &$\le8.26\times 10^{-9}$   &$\le3.00\times 10^{-10}$   &$\le 2.58$   &$\le-2.92$\\
                    S &$\,^{(a)}$   &$\le-2.93$   &$\le3.64\times 10^{-7}$   &$\le1.27\times 10^{-8}$   &$\le 4.21$   &$\le-3.00$\\
                    Ar&$\,^{(a)}$   &   $ 1.07$   &   $8.65\times 10^{-4}$   &   $2.43\times 10^{-5}$   &   $ 7.49$   &   $ 1.00$\\
                    Ca&$\,^{(a)}$   &$\le 0.02$   &$\le6.41\times 10^{-5}$   &$\le1.80\times 10^{-6}$   &$\le 6.35$   &$\le-0.07$\\
                    Sc&$\,^{(a)}$   &$\le-0.01$   &$\le4.64\times 10^{-8}$   &$\le1.16\times 10^{-9}$   &$\le 3.16$   &$\le-0.07$\\
                    Ti&$\,^{(a)}$   &$\le 0.02$   &$\le3.12\times 10^{-6}$   &$\le7.31\times 10^{-8}$   &$\le 4.96$   &$\le-0.07$\\
                    V &$\,^{(a)}$   &$\le 0.04$   &$\le3.17\times 10^{-7}$   &$\le6.99\times 10^{-9}$   &$\le 3.94$   &$\le-0.07$\\
                    Cr&$\,^{(a)}$   &   $ 1.86$   &   $1.16\times 10^{-3}$   &   $2.50\times 10^{-5}$   &   $ 7.50$   &   $ 1.77$\\
                    Mn&$\,^{(a)}$   &   $ 1.56$   &   $3.81\times 10^{-4}$   &   $7.80\times 10^{-6}$   &   $ 6.99$   &   $ 1.48$\\
                    Fe&$\,^{(a)}$   &   $ 0.10$   &   $1.52\times 10^{-3}$   &   $3.06\times 10^{-5}$   &   $ 7.58$   &   $ 0.00$\\
                    Co&$\,^{(a)}$   &   $ 2.18$   &   $1.57\times 10^{-4}$   &   $3.00\times 10^{-6}$   &   $ 6.58$   &   $ 1.50$\\
                    Ni&$\,^{(a)}$   &   $ 0.70$   &   $3.39\times 10^{-4}$   &   $6.49\times 10^{-6}$   &   $ 6.91$   &   $ 0.61$\\
                    Cu&             &$\le 1.38$   &$\le1.70\times 10^{-5}$   &$\le3.00\times 10^{-7}$   &$\le 5.58$   &$\le 1.30$\\
                    Zn&             &   $ 1.37$   &   $4.08\times 10^{-5}$   &   $7.00\times 10^{-7}$   &   $ 5.95$   &   $ 1.30$\\
                    Ga&             &   $ 3.77$   &   $3.11\times 10^{-4}$   &   $5.00\times 10^{-6}$   &   $ 6.80$   &   $ 3.67$\\
                    Ge&             &   $ 3.76$   &   $1.29\times 10^{-3}$   &   $2.00\times 10^{-5}$   &   $ 7.40$   &   $ 3.67$\\
                    Se&             &   $ 4.52$   &   $4.21\times 10^{-3}$   &   $6.00\times 10^{-5}$   &   $ 7.88$   &   $ 4.45$\\
                    Br&             &   $ 3.19$   &   $3.13\times 10^{-5}$   &   $4.40\times 10^{-7}$   &   $ 5.74$   &   $ 3.12$\\
                    Kr&             &   $ 2.68$   &   $5.23\times 10^{-5}$   &   $7.00\times 10^{-7}$   &   $ 5.95$   &   $ 2.61$\\
                    Sr&             &   $ 5.21$   &   $6.98\times 10^{-3}$   &   $9.00\times 10^{-5}$   &   $ 8.05$   &   $ 5.10$\\
                    Zr&             &   $ 3.97$   &   $2.44\times 10^{-4}$   &   $3.00\times 10^{-6}$   &   $ 6.58$   &   $ 3.91$\\
                    Mo&             &   $ 3.68$   &   $2.57\times 10^{-5}$   &   $3.00\times 10^{-7}$   &   $ 5.58$   &   $ 3.61$\\
                    In&             &   $ 5.83$   &   $3.58\times 10^{-4}$   &   $3.50\times 10^{-6}$   &   $ 6.64$   &   $ 5.76$\\
                    Te&             &   $ 4.51$   &   $4.55\times 10^{-4}$   &   $4.00\times 10^{-6}$   &   $ 6.70$   &   $ 4.44$\\
                    I &             &   $ 4.24$   &   $5.66\times 10^{-5}$   &   $5.00\times 10^{-7}$   &   $ 5.80$   &   $ 4.16$\\
                    Xe&             &   $ 3.85$   &   $1.17\times 10^{-4}$   &   $1.00\times 10^{-6}$   &   $ 6.10$   &   $ 3.77$\\
                    Ba&             &   $ 3.54$   &   $6.12\times 10^{-5}$   &   $5.00\times 10^{-7}$   &   $ 5.80$   &   $ 3.53$\\
\hline
\end{tabular}
\newline
\textbf{Notes.}
$^{(a)}$From \citet{ziegleretal2012}.
$^{(b)}$Interpolated from post-AGB evolutionary tracks, cf\@., Fig.\,\ref{fig:loggteff}.
$^{(c)}$Interpolated from post-EHB evolutionary tracks, cf\@., Fig.\,\ref{fig:loggteff}.
$^{(d)}$Abundances $\varepsilon_i = \log n_i + c$ with $\sum_i a_i n_i = 12.15$ and the atomic weights $a_i$.
\end{table*}

\clearpage

\section{Additional figures and tables.}
\label{app:additional}

\begin{table*}
\begin{center}
\centering
\caption{Observation log for \bd.}
\label{tab:obslog}
\vspace*{-2mm}
\begin{tabular}{llcrrrr}
\hline
\hline
  &  &  & \multicolumn{1}{c}{Wavelength}  & \multicolumn{1}{c}{Aperture/} & \multicolumn{1}{c}{Exposure}  & \multicolumn{1}{c}{Resolving power}\\
\vspace{-0.52cm}\\
 \multicolumn{1}{c}{Instrument} & \multicolumn{1}{c}{Dataset Id} & \multicolumn{1}{c}{Start Time (UT)} &  &  &  &\\
\vspace{-0.52cm}\\
  &  & & \multicolumn{1}{c}{range ($\lambda$)} & \multicolumn{1}{c}{Grating} & \multicolumn{1}{c}{time (s)} &\multicolumn{1}{c}{$R = \lambda\,/\,\Delta\lambda$}\\
\hline
 FUSE\,$^\mathrm{a}$ & P1330101000 & 2000-05-20 20:27:37 & $910 - 1180$ & LWRS & $4416$ & {20\,000}\\
 STIS\,$^\mathrm{b}$  & O4GT02010   & 1999-04-17 21:14:49 & $1150-1730$ & E140M & $2050$ & 45\,800\\
 STIS  & O4GT02020   & 1999-04-17 22:37:03 & $1150-1730$ & E140M & $2800$ & 45\,800\\
 STIS  & O4GT02030   & 1999-04-18 00:16:10 & $1150-1730$ & E140M & $2740$ & 45\,800\\
 STIS  & O4GT02040   & 1999-04-18 01:52:54 & $1150-1730$ & E140M & $2740$ & 45\,800\\
\hline
\end{tabular}
\end{center}
\vspace{-3mm}
\begin{footnotesize}
\raggedright{
\textbf{Notes.} 
a: \hspace{0.5mm} Far Ultraviolet Spectroscopic Explorer,\hspace{1.5mm}
b: \hspace{0.5mm} Space Telescope Imaging Spectrograph.\hspace{1.5mm}\\
}
\end{footnotesize}
\end{table*}

\begin{figure*}
  \resizebox{\hsize}{!}{\includegraphics{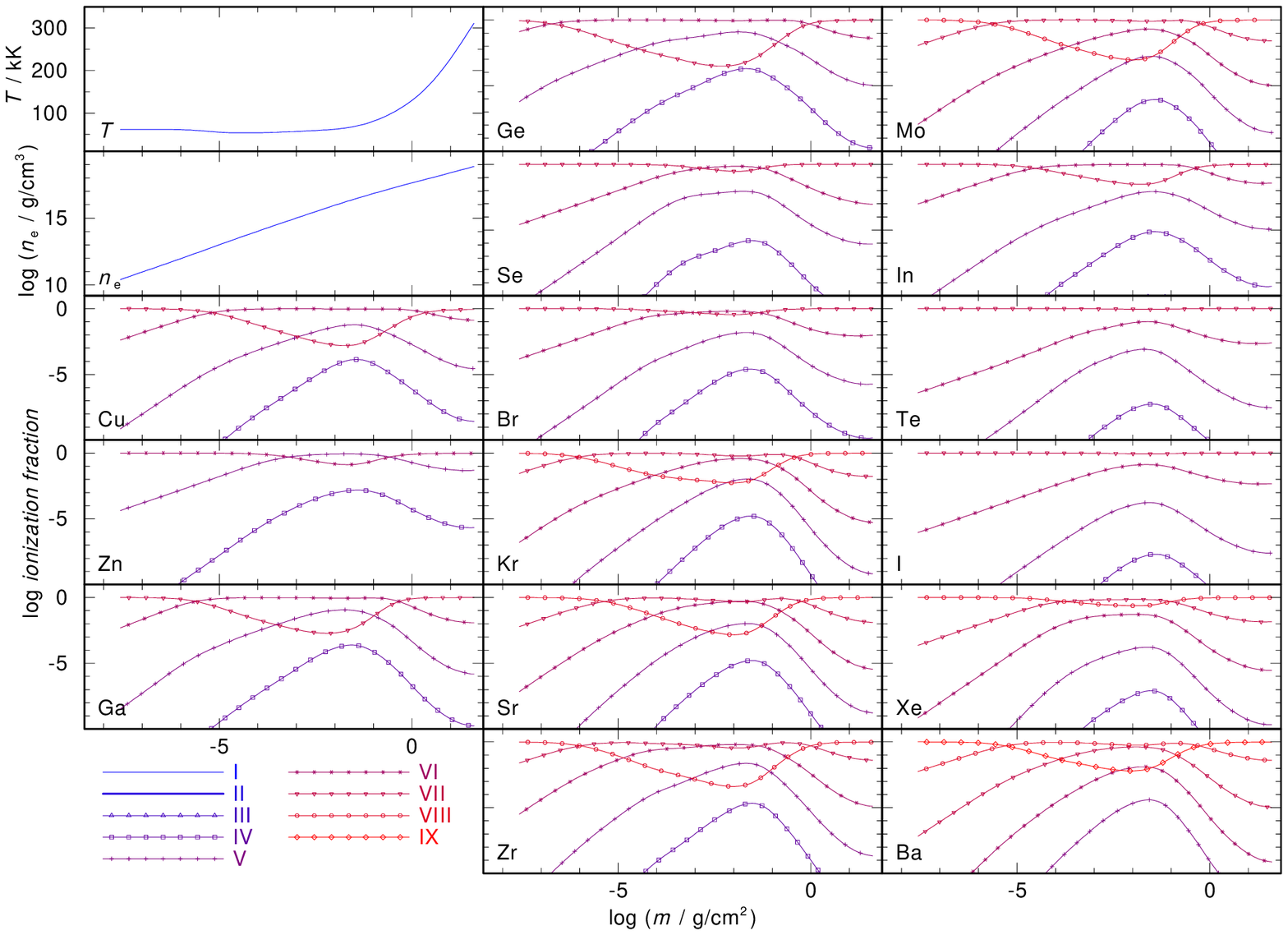}} 
   \caption[]{Temperature and electron density structure and ionization fractions of all TIE ions which are considered in our final model for \bd.
             } 
   \label{fig:ionfrac}
\end{figure*}

\begin{footnotesize}
\begin{table*}\fontsize{8.2}{8.5}\selectfont
\centering
\caption{Statistics of the H -- Ar$^{a}$ and Ca - Ba$^{b}$ model atoms used in our model-atmosphere calculations.}
\label{tab:stat}
\setlength{\tabcolsep}{.3em}
\begin{tabular}{rlrrrp{6mm}rlrrrp{6mm}rlrrr}
\hline
\hline
\multicolumn{2}{l}{}    & \multicolumn{2}{c}{Levels} & & &\multicolumn{2}{l}{}& \multicolumn{1}{l}{Super}                   & \multicolumn{1}{c}{Super} & Individual & &\multicolumn{2}{l}{}& \multicolumn{1}{l}{Super}                   & \multicolumn{1}{c}{Super} & Individual\\
\cline{3-4}
\multicolumn{2}{l}{}    &      &        &          && \multicolumn{2}{l}{}    &                         &        &       \vspace{-4.0mm}\\
\multicolumn{2}{l}{\hbox{}\hspace{4mm}Ion} &      &        &  ~Lines  && 
\multicolumn{2}{l}{\hbox{}\hspace{4mm}Ion} &        &        &       & 
\multicolumn{2}{l}{\hbox{}\hspace{4mm}Ion} &        &        &       \vspace{-1.0mm}\\
\multicolumn{2}{l}{}    & NLTE & ~~~LTE &          && \multicolumn{2}{l}{}    & \multicolumn{1}{c}{levels$^{c}$} & \multicolumn{1}{c}{lines}  & \multicolumn{1}{c}{lines} && \multicolumn{2}{l}{}    & \multicolumn{1}{c}{levels$^{c}$} & \multicolumn{1}{c}{lines}  & \multicolumn{1}{c}{lines}               \\
\hline
H  & \Ion{}{1} &  12 &      4  &      66    & & Ca & \Ion{}{4} &   6 &     16  &      20291  & &        Cu & \Ion{}{4}      & 7   &     15  &       8785\\
   & \Ion{}{2} &   1 &    $-$  &     $-$    & &    & \Ion{}{5} &   6 &     21  &     141956  & & 	   & \Ion{}{5}      & 7   &     16  &       5456\\
He & \Ion{}{1} &   5 &     98  &       3    & &    & \Ion{}{6} &   6 &     19  &     114545  & & 	   & \Ion{}{6}      & 7   &      9  &       3797\\
   & \Ion{}{2} &  16 &     16  &     120    & &    & \Ion{}{7} &   6 &     21  &      71608  & & 	   & \Ion{}{7}      & 1   &      0  &          0\\
   & \Ion{}{3} &   1 &    $-$  &     $-$    & &    & \Ion{}{8} &   6 &     20  &       9124  & & 	Zn & \Ion{}{4}      &   7 &     11  &        400\\         
C  & \Ion{}{3} &   6 &     61  &      12    & &    & \Ion{}{9} &   1 &      0  &          0  & & 	   & \Ion{}{5}      &   7 &     15  &       1879\\         
   & \Ion{}{4} &  54 &      4  &     295    & & Sc & \Ion{}{4} &   6 &     20  &      15024  & & 	   & \Ion{}{6}      &   1 &      0  &          0\\         
   & \Ion{}{5} &   1 &      0  &       0    & &    & \Ion{}{5} &   6 &     21  &     261235  & & 	Ga & \Ion{}{4}      &   7 &     19  &       3198\\         
N  & \Ion{}{3} &   1 &     65  &       0    & &    & \Ion{}{6} &   6 &     19  &     237271  & & 	   & \Ion{}{5}      &   7 &     15  &        517\\         
   & \Ion{}{4} &  16 &     78  &      30    & &    & \Ion{}{7} &   6 &     20  &     176143  & & 	   & \Ion{}{6}      &   7 &     13  &       1914\\         
   & \Ion{}{5} &  54 &      8  &     297    & &    & \Ion{}{8} &   6 &     21  &      91935  & & 	   & \Ion{}{7}      &   1 &      0  &          0\\         
   & \Ion{}{6} &   1 &      0  &       0    & &    & \Ion{}{9} &   1 &      0  &          0  & & 	Ge & \Ion{}{4}$^{d}$ &   8 &     0  &          8\\         
O  & \Ion{}{3} &   3 &     69  &       0    & & Ti & \Ion{}{4} &   6 &     19  &       1000  & & 	   & \Ion{}{5} &   7 &     16  &       2159\\              
   & \Ion{}{4} &  18 &     76  &      39    & &    & \Ion{}{5} &   6 &     20  &      26654  & & 	   & \Ion{}{6} &   7 &     12  &        414\\              
   & \Ion{}{5} &  90 &     36  &     610    & &    & \Ion{}{6} &   6 &     19  &      95448  & & 	   & \Ion{}{7} &   1 &      0  &          0\\              
   & \Ion{}{6} &  54 &      8  &     291    & &    & \Ion{}{7} &   6 &     20  &     230618  & & 	Se & \Ion{}{4} &   1 &      0  &          0\\              
   & \Ion{}{7} &   1 &      0  &       0    & &    & \Ion{}{8} &   6 &     21  &     182699  & & 	   & \Ion{}{5} &   7 &     19  &        310\\              
F  & \Ion{}{3} &   1 &      6  &       0    & &    & \Ion{}{9} &   1 &      0  &          0  & & 	   & \Ion{}{6} &   1 &      0  &          0\\              
   & \Ion{}{4} &   1 &     10  &       0    & & V  & \Ion{}{4} &   6 &     19  &      37130  & & 	   & \Ion{}{7} &   1 &      0  &          0\\              
   & \Ion{}{5} &  15 &     91  &      31    & &    & \Ion{}{5} &   6 &     20  &       2123  & & 	Br & \Ion{}{3} &   1 &      0  &          0\\              
   & \Ion{}{6} &  12 &    115  &      16    & &    & \Ion{}{6} &   6 &     19  &      35251  & & 	   & \Ion{}{4} &   6 &     12  &        424\\              
   & \Ion{}{7} &   1 &      0  &       0    & &    & \Ion{}{7} &   6 &     19  &     112883  & & 	   & \Ion{}{5} &   7 &     18  &        394\\              
Ne & \Ion{}{2} &   1 &     33  &       0    & &    & \Ion{}{8} &   6 &     20  &     345089  & & 	   & \Ion{}{6} &   7 &     17  &        158\\              
   & \Ion{}{3} &   3 &     43  &       0    & &    & \Ion{}{9} &   1 &      0  &          0  & & 	   & \Ion{}{7} &   1 &      0  &          0\\              
   & \Ion{}{4} &   3 &     37  &       0    & & Cr & \Ion{}{4} &   6 &     20  &     234170  & & 	Kr & \Ion{}{4} &   7 &     19  &        911\\              
   & \Ion{}{5} &  20 &     74  &      35    & &    & \Ion{}{5} &   6 &     20  &      43860  & & 	   & \Ion{}{5} &   7 &     16  &        553\\              
   & \Ion{}{6} &   1 &      0  &       0    & &    & \Ion{}{6} &   6 &     20  &       4406  & & 	   & \Ion{}{6} &   7 &     19  &        843\\              
Na & \Ion{}{3} &   1 &    186  &       0    & &    & \Ion{}{7} &   6 &     19  &      37070  & & 	   & \Ion{}{7} &   7 &     21  &        743\\              
   & \Ion{}{4} &   1 &    237  &       0    & &    & \Ion{}{8} &   6 &     20  &     132221  & & 	   & \Ion{}{8} &   1 &      0  &          0\\              
   & \Ion{}{5} &   8 &     42  &       9    & &    & \Ion{}{9} &   1 &      0  &          0  & & 	Sr & \Ion{}{4} &   7 &     21  &       7578\\              
   & \Ion{}{6} &  43 &     10  &     130    & & Mn & \Ion{}{4} &   6 &     20  &     719387  & & 	   & \Ion{}{5} &   7 &     19  &       2022\\              
   & \Ion{}{7} &   1 &      0  &       0    & &    & \Ion{}{5} &   6 &     20  &     285376  & & 	   & \Ion{}{6} &   7 &     10  &         70\\              
Mg & \Ion{}{3} &   1 &     34  &       0    & &    & \Ion{}{6} &   6 &     20  &      70116  & & 	   & \Ion{}{7} &   7 &     10  &         46\\              
   & \Ion{}{4} &  31 &      0  &      93    & &    & \Ion{}{7} &   6 &     20  &       8277  & & 	   & \Ion{}{8} &   1 &      0  &          0\\              
   & \Ion{}{5} &  15 &     37  &      18    & &    & \Ion{}{8} &   6 &     20  &      37168  & & 	Zr & \Ion{}{4} &   7 &     20  &        135\\              
   & \Ion{}{6} &   1 &      0  &       0    & &    & \Ion{}{9} &   1 &      0  &          0  & & 	   & \Ion{}{5} &   7 &     22  &       1449\\              
Al & \Ion{}{3} &   1 &      6  &       0    & & Fe & \Ion{}{4} &   6 &     20  &    3102371  & & 	   & \Ion{}{6} &   7 &     12  &       1098\\              
   & \Ion{}{4} &  15 &      2  &       0    & &    & \Ion{}{5} &   6 &     20  &    3266247  & & 	   & \Ion{}{7} &   7 &     15  &        947\\              
   & \Ion{}{5} &   1 &     16  &       0    & &    & \Ion{}{6} &   6 &     20  &     991935  & & 	   & \Ion{}{8} &   1 &      0  &          0\\              
   & \Ion{}{6} &  14 &     24  &      16    & &    & \Ion{}{7} &   6 &     20  &     200455  & & 	Mo & \Ion{}{4} &   7 &     15  &       2803\\              
   & \Ion{}{7} &   1 &      0  &       0    & &    & \Ion{}{8} &   6 &     18  &      19587  & & 	   & \Ion{}{5} &   7 &     22  &       5829\\              
Si & \Ion{}{3} &   3 &     31  &       1    & &    & \Ion{}{9} &   1 &      0  &          0  & & 	   & \Ion{}{6} &   7 &     23  &        984\\              
   & \Ion{}{4} &  16 &      7  &      44    & & Co & \Ion{}{4} &   6 &     20  &     552916  & & 	   & \Ion{}{7} &   7 &     16  &       1173\\              
   & \Ion{}{5} &  25 &      0  &      59    & &    & \Ion{}{5} &   6 &     20  &    1469717  & & 	   & \Ion{}{8} &   1 &      0  &          0\\              
   & \Ion{}{6} &   1 &      0  &       0    & &    & \Ion{}{6} &   6 &     18  &     898484  & & 	In & \Ion{}{3} &   1 &      0  &          0\\              
P  & \Ion{}{3} &   1 &      9  &       0    & &    & \Ion{}{7} &   6 &     19  &     492913  & & 	   & \Ion{}{4} &   7 &     14  &        564\\              
   & \Ion{}{4} &  15 &     36  &       9    & &    & \Ion{}{8} &   6 &     20  &      88548  & & 	   & \Ion{}{5} &   7 &     10  &        919\\              
   & \Ion{}{5} &  18 &      7  &      12    & &    & \Ion{}{9} &   1 &      0  &          0  & & 	   & \Ion{}{6} &   8 &     10  &        176\\              
   & \Ion{}{6} &   1 &      0  &       0    & & Ni & \Ion{}{4} &   6 &     20  &    2512561  & & 	   & \Ion{}{7} &   1 &      0  &          0\\              
S  & \Ion{}{4} &   6 &     94  &       4    & &    & \Ion{}{5} &   6 &     20  &    2766664  & & 	Te & \Ion{}{4} &   1 &      0  &          0\\              
   & \Ion{}{5} &  21 &     89  &      37    & &    & \Ion{}{6} &   6 &     18  &    7408657  & & 	   & \Ion{}{5} &   1 &      0  &          0\\              
   & \Ion{}{6} &  18 &     19  &      48    & &    & \Ion{}{7} &   6 &     18  &    4195381  & & 	   & \Ion{}{6} &   7 &     12  &        178\\              
   & \Ion{}{7} &   1 &      0  &       0    & &    & \Ion{}{8} &   6 &     20  &    1473122  & & 	   & \Ion{}{7} &   1 &      0  &          0\\              
Ar & \Ion{}{4} &   1 &    349  &       0    & &    & \Ion{}{9} &   1 &      0  &          0  & & 	I  & \Ion{}{4} &   1 &      0  &          0\\              
   & \Ion{}{5} &  32 &    329  &      38    & &    &           &     &         &             & & 	   & \Ion{}{5} &   1 &      0  &          0\\              
   & \Ion{}{6} &  16 &    168  &      21    & &    &           &     &         &             & & 	   & \Ion{}{6} &   7 &     15  &        197\\              
   & \Ion{}{7} &  40 &    112  &     130    & &    &           &     &         &             & & 	   & \Ion{}{7} &   1 &      0  &          0\\              
   & \Ion{}{8} &   1 &      0  &       0    & &    &           &     &         &             & & 	Xe & \Ion{}{4} &   7 &     16  &       1391\\              
   &           &     &         &            & &    &           &     &         &             & &  	   & \Ion{}{5} &   7 &     15  &        616\\              
   &           &     &         &            & &    &           &     &         &             & &  	   & \Ion{}{6} &   7 &     16  &        243\\              
   &           &     &         &            & &    &           &     &         &             & &   	   & \Ion{}{7} &   7 &     19  &        491\\              
   &           &     &         &            & &    &           &     &         &             & &   	   & \Ion{}{8} &   1 &      0  &          0\\              
   &           &     &         &            & &    &           &     &         &             & &   	Ba & \Ion{}{5} &   7 &     12  &        981\\              
   &           &     &         &            & &    &           &     &         &             & &   	   & \Ion{}{6} &   7 &      6  &        162\\              
   &           &     &         &            & &    &           &     &         &             & &   	   & \Ion{}{7} &   7 &     11  &        493\\              
   &           &     &         &            & &    &           &     &         &             & &   	   & \Ion{}{8}$^{e}$ &   34 &      0  &         44\\       
   &           &     &         &            & &    &           &     &         &             & &   	   & \Ion{}{9} &   1 &      0  &          0\\              
\noalign{\smallskip}
\cline{1-17}
\noalign{\smallskip}
total &        & 742 &    2776 &       2514 & &    &           & 279 &   884   &  33219636   & &           &           & 359 &     651 &   63444\\
\hline
\end{tabular}

\textbf{Notes.} $^{(a)}${classical model atoms},
$^{(b)}${model atoms constructed using a statistical approach \citep{rauchdeetjen2003}},\\
$^{(c)}${treated as NLTE levels}, $^{(d)}${\Ion{Ge}{4} classical model atom with 8 NLTE levels, 1 LTE level, and 8 transitions},\\
$^{(e)}${\Ion{Ba}{8} classical model atom with 34 NLTE levels, 0 LTE level, and 44 transitions}.
\end{table*}
\end{footnotesize}

\begin{figure*}
  \resizebox{\hsize}{!}{\includegraphics{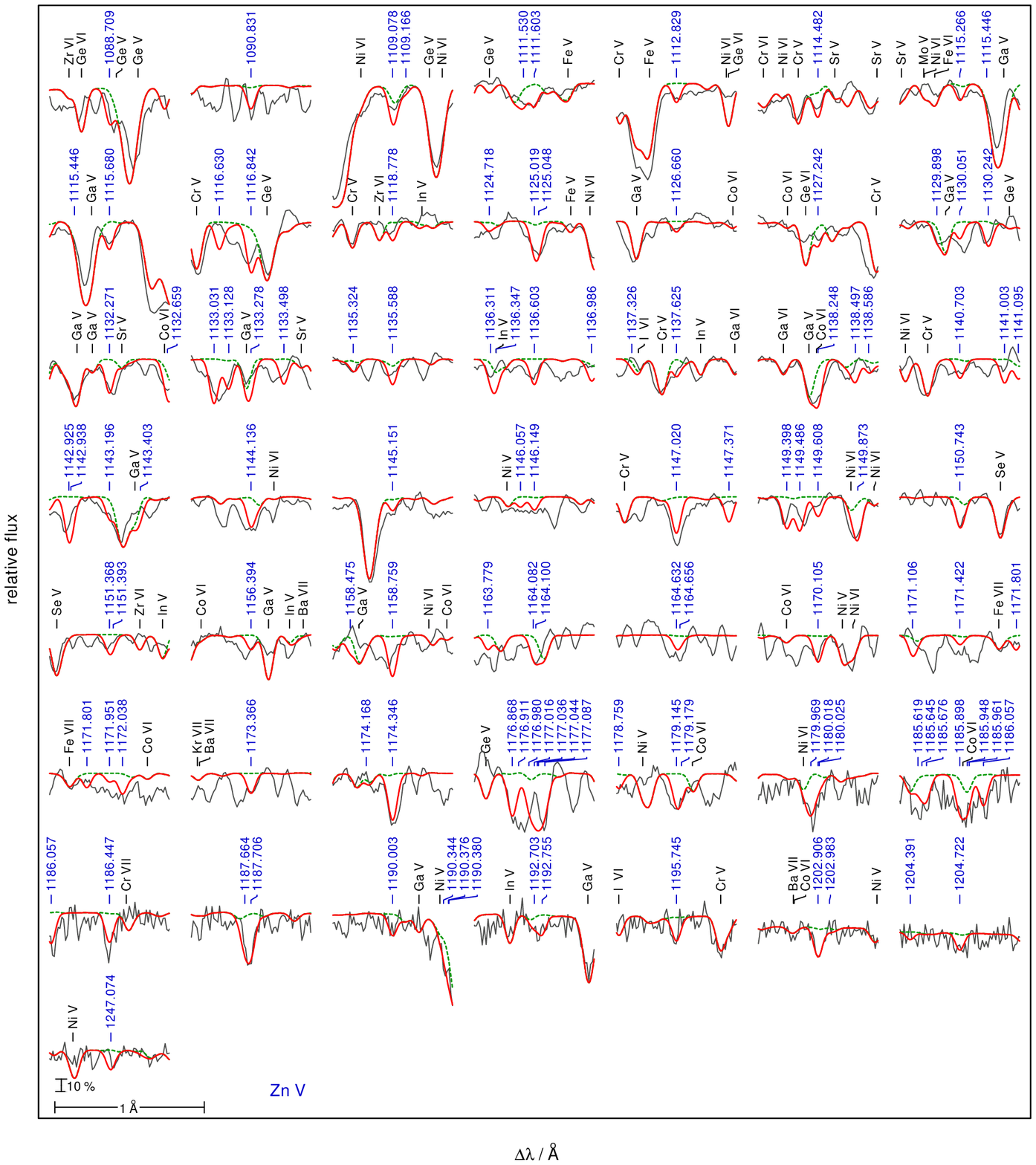}}
   \caption{Prominent computed and identified lines of \Ion{Zn}{5} (blue wavelength labels) in the FUSE ($\lambda < 1180$\,{\AA}) and STIS observations of \bd. The model was calculated with the abundances given in \ta{tab:finab} (red). In addition, a model without Zn (green dashed) is shown.}
\label{fig:zn}
\end{figure*}

\begin{figure*}
  \resizebox{\hsize}{!}{\includegraphics{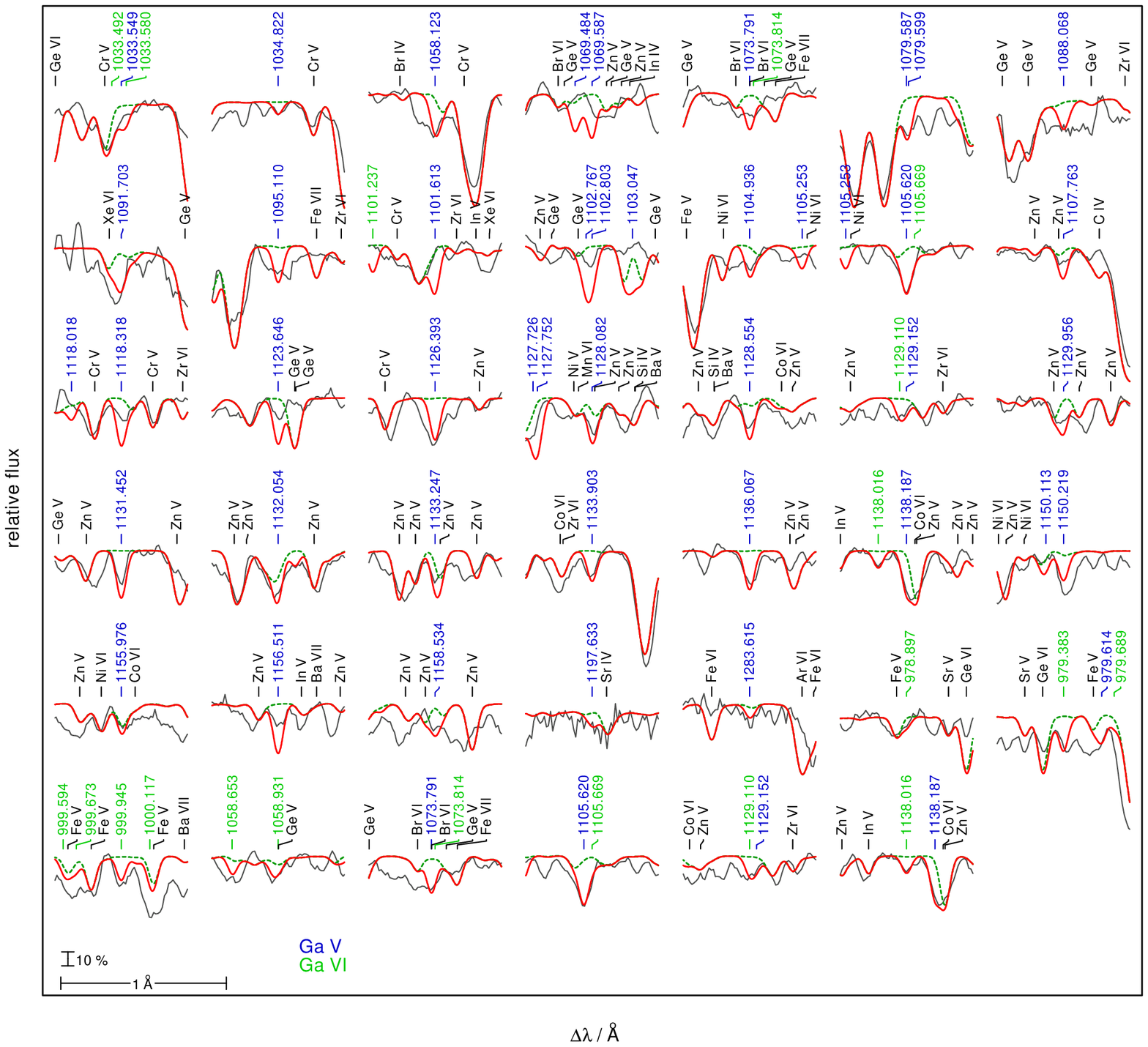}}
   \caption{Like \ab{fig:zn}, for \Ion{Ga}{5} (blue) and \Ion{Ga}{6} (green).}
\label{fig:ga}
\end{figure*}

\begin{figure*}
  \resizebox{\hsize}{!}{\includegraphics{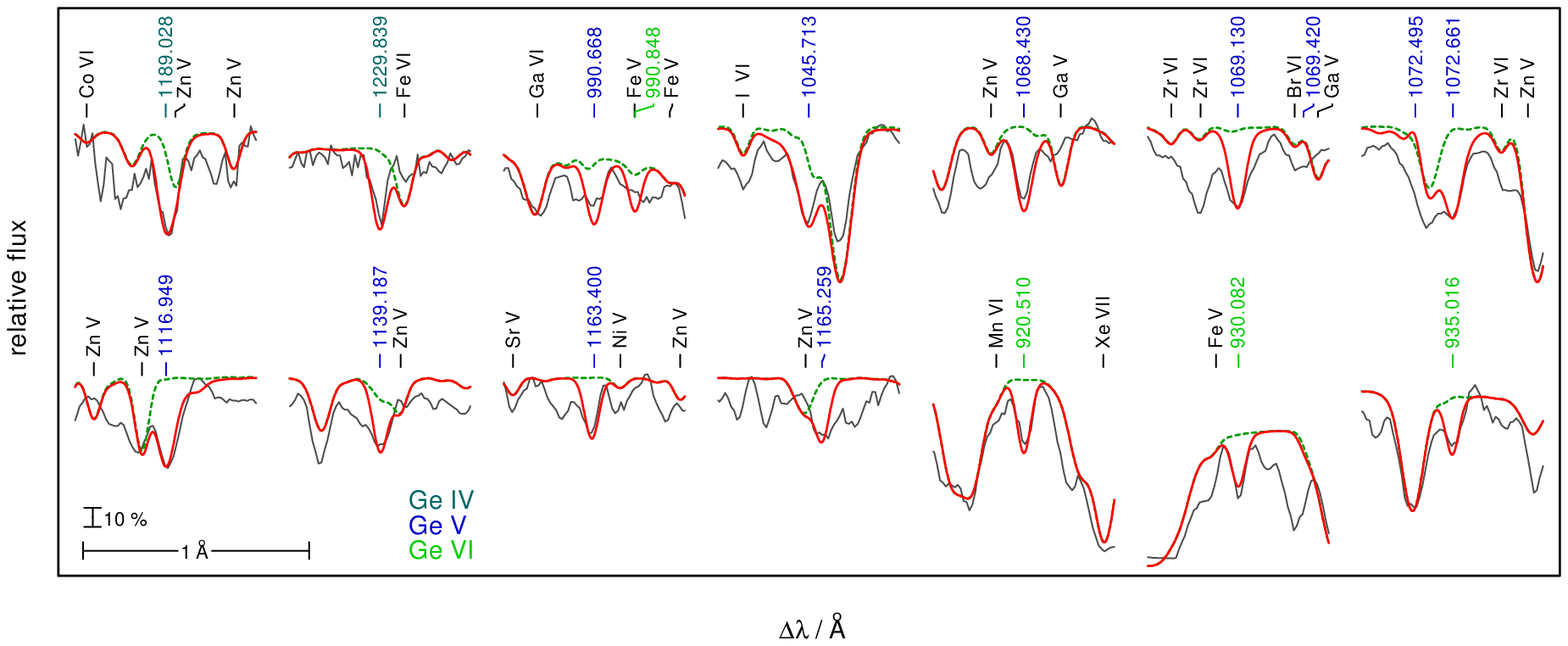}}
   \caption{Like \ab{fig:zn}, for \Ion{Ge}{4} ({dark cyan}), \Ion{Ge}{5} (blue), and \Ion{Ge}{6} (green).}
\label{fig:ge}
\end{figure*}

\begin{figure*}
  \resizebox{\hsize}{!}{\includegraphics{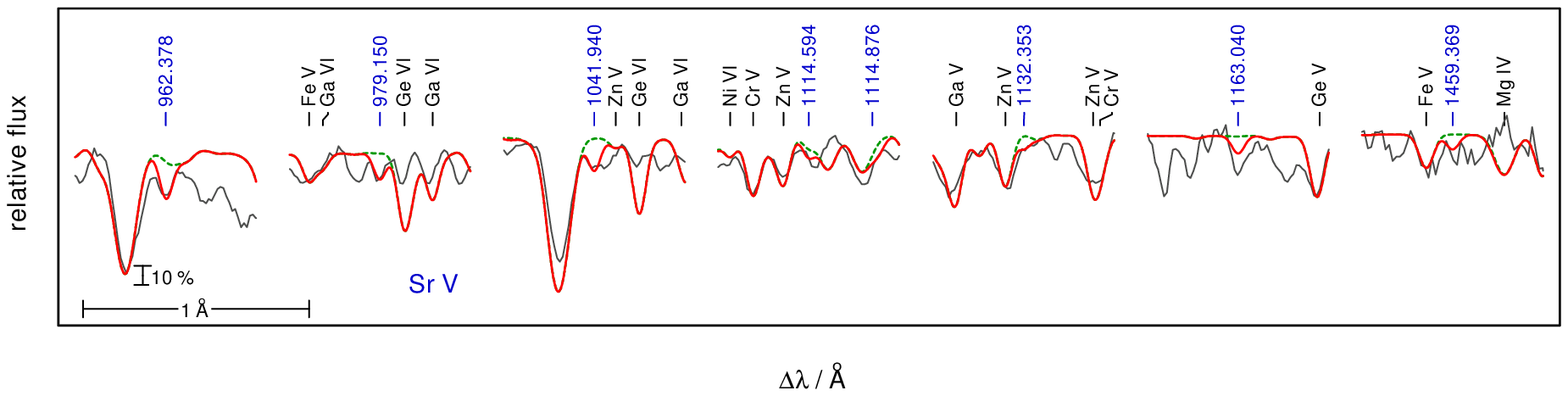}}
   \caption{Like \ab{fig:zn}, for \Ion{Sr}{5} (blue).}
\label{fig:sr}
\end{figure*}

\begin{figure*}
  \resizebox{\hsize}{!}{\includegraphics{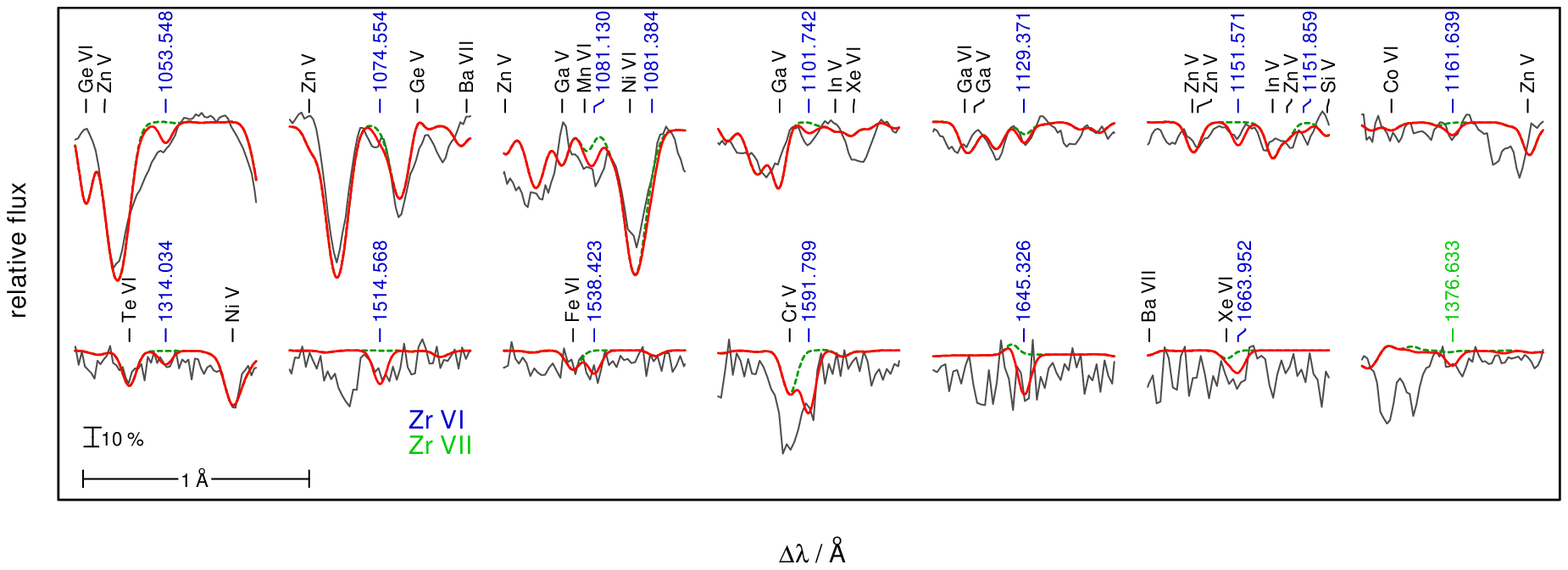}}
   \caption{Like \ab{fig:zn}, for \Ion{Zr}{6} (blue) and \Ion{Zr}{7} (green).}
\label{fig:zr}
\end{figure*}

\begin{figure*}
  \resizebox{\hsize}{!}{\includegraphics{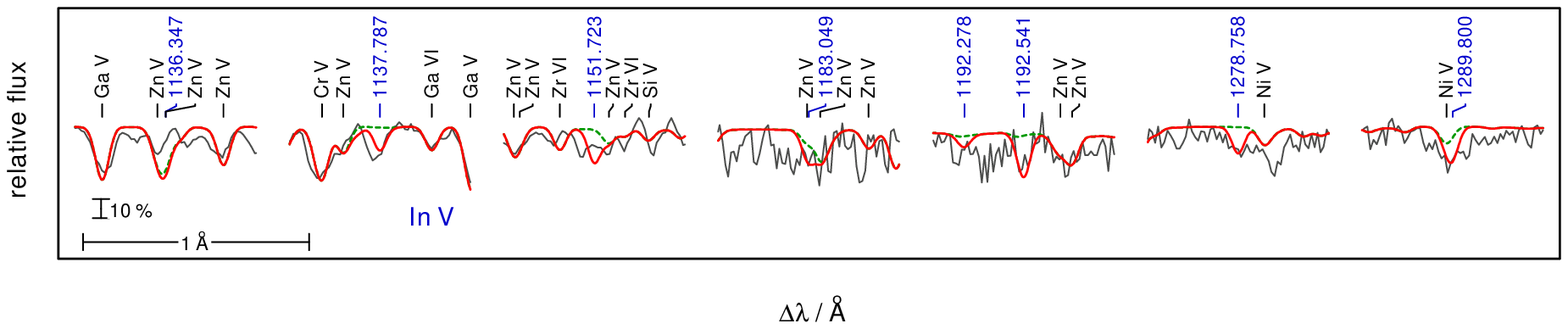}}
   \caption{Like \ab{fig:zn}, for \Ion{In}{5} (blue).}
\label{fig:in}
\end{figure*}

\begin{figure*}
  \resizebox{\hsize}{!}{\includegraphics{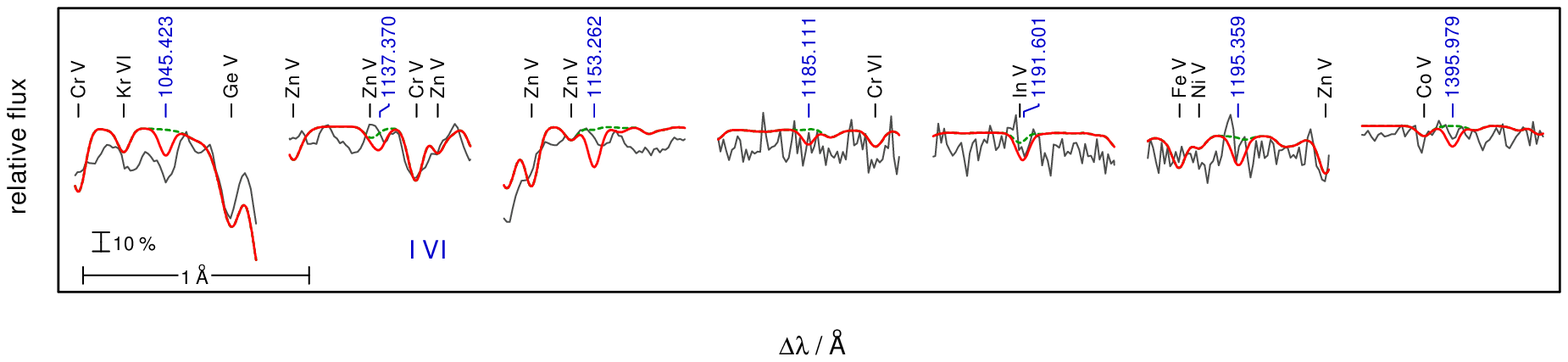}}
   \caption{Like \ab{fig:zn}, for \Ion{I}{6} (blue).}
\label{fig:i}
\end{figure*}
\begin{figure}
  \resizebox{\hsize}{!}{\includegraphics{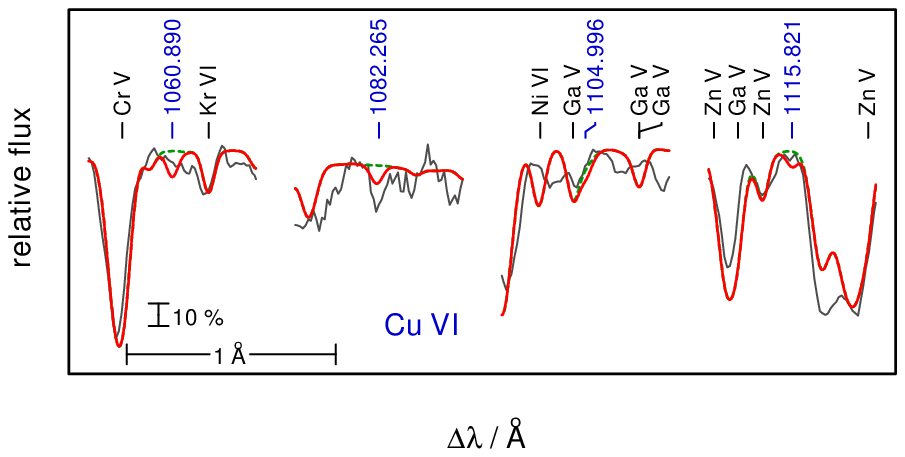}}
   \caption{Like \ab{fig:zn}, for \Ion{Cu}{6} (blue).}
\label{fig:cu}
\end{figure}

\begin{figure}
  \resizebox{\hsize}{!}{\includegraphics{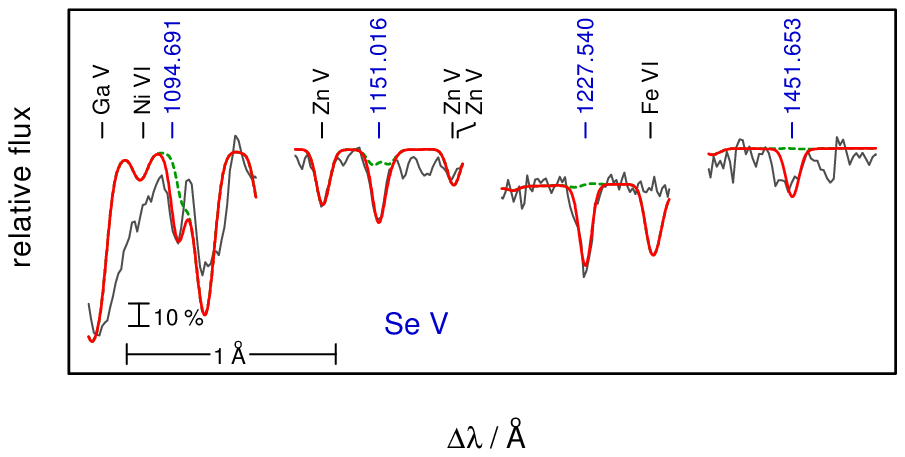}}
   \caption{Like \ab{fig:zn}, for \Ion{Se}{5} (blue).}
\label{fig:se}
\end{figure}

\begin{figure}
  \resizebox{\hsize}{!}{\includegraphics{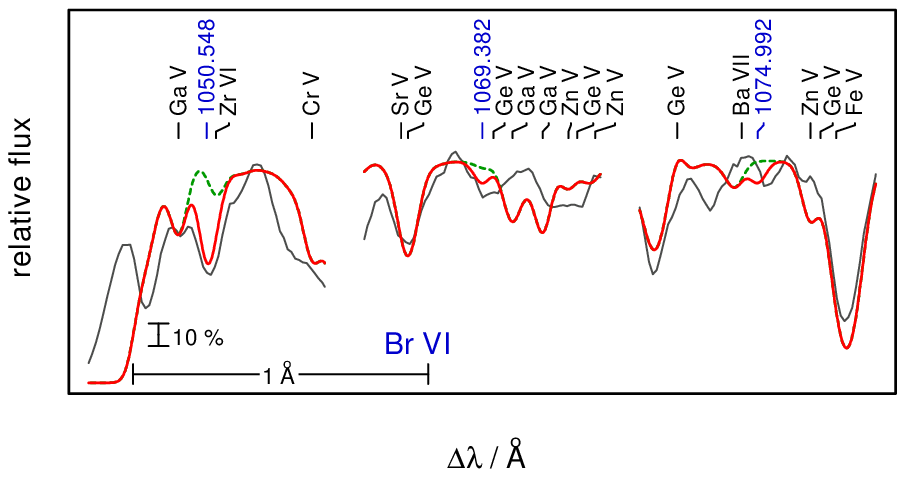}}
   \caption{Like \ab{fig:zn}, for \Ion{Br}{6} (blue).}
\label{fig:br}
\end{figure}

\begin{figure}
  \resizebox{\hsize}{!}{\includegraphics{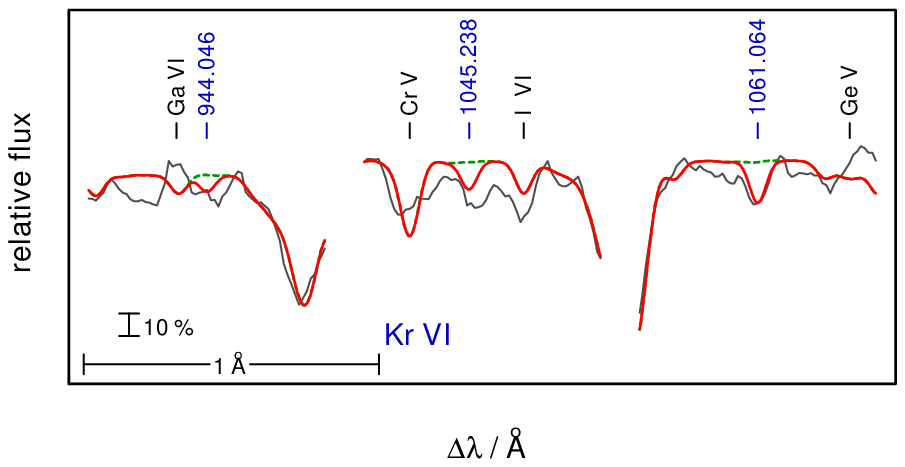}}
   \caption{Like \ab{fig:zn}, for \Ion{Kr}{6} (blue).}
\label{fig:kr}
\end{figure}

\begin{figure}
  \resizebox{\hsize}{!}{\includegraphics{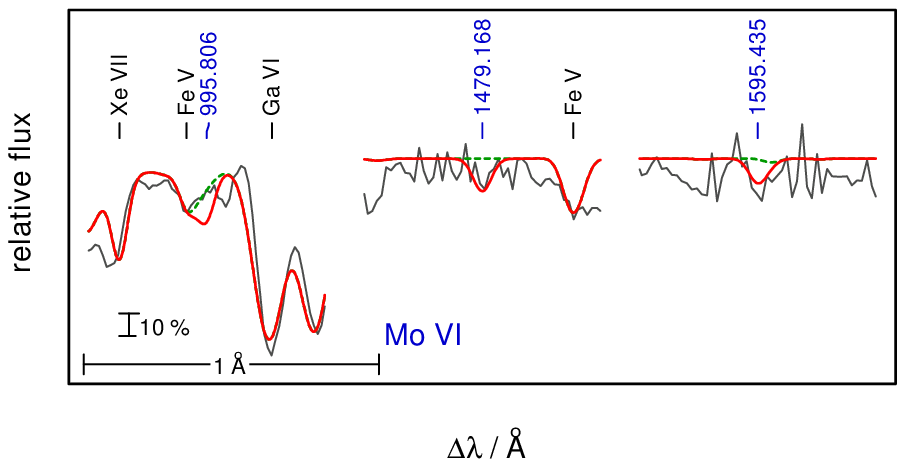}}
   \caption{Like \ab{fig:zn}, for \Ion{Mo}{6} (blue).}
\label{fig:mo}
\end{figure}

\begin{figure}
  \resizebox{\hsize}{!}{\includegraphics{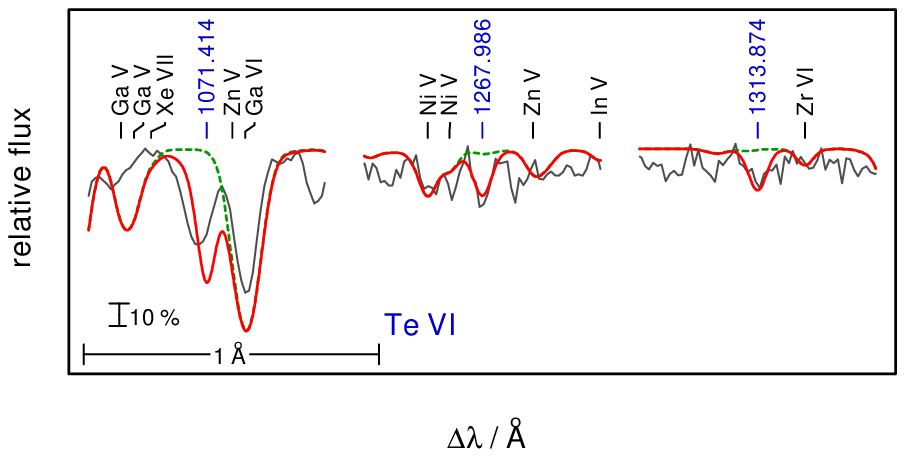}}
   \caption{Like \ab{fig:zn}, for \Ion{Te}{6} (blue).}
\label{fig:te}
\end{figure}

\begin{figure}
  \resizebox{\hsize}{!}{\includegraphics{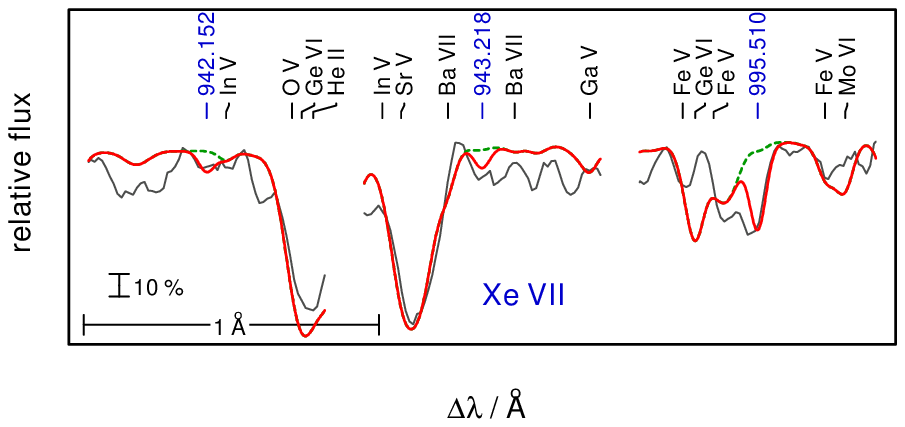}}
   \caption{Like \ab{fig:zn}, for \Ion{Xe}{7} (blue).}
\label{fig:xe}
\end{figure}

\begin{figure}
  \resizebox{\hsize}{!}{\includegraphics{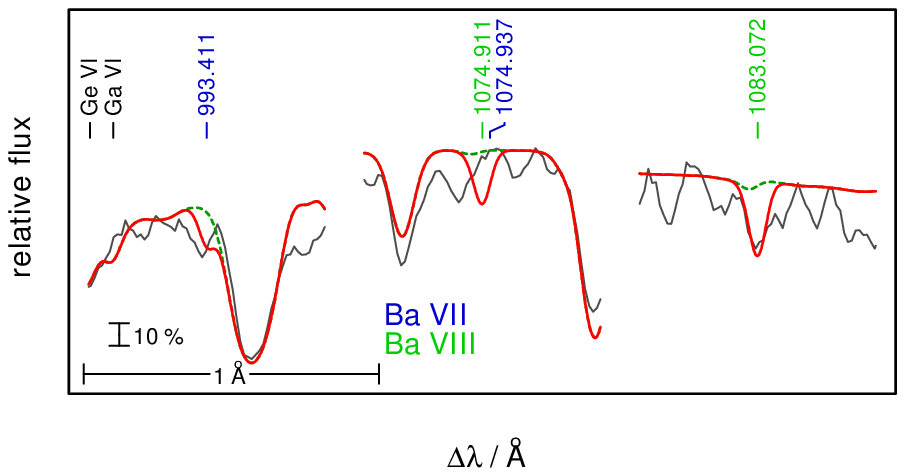}}
   \caption{Like \ab{fig:zn}, for \Ion{Ba}{7} (blue) and \Ion{Ba}{8} (green).}
\label{fig:ba}
\end{figure}

\onecolumn

\begin{longtable}{rlp{5.5cm}l}
\caption{\label{tab:Culines} Cu lines with $W_\lambda \ge 5$\,m{\AA} in the model spectrum of \bd.} \\
\hline\hline
\noalign{\smallskip}
Ion & Stage & Wavelength\,/\,{\AA} & Comment \\ 
\noalign{\smallskip}
\hline
\endfirsthead
\caption{continued.}\\
\hline\hline
\noalign{\smallskip}
Ion & Stage & Wavelength\,/\,{\AA} & Comment \\
\noalign{\smallskip}
\hline
\noalign{\smallskip}
\endhead
\hline
\noalign{\smallskip}
\endfoot
\noalign{\smallskip}
Cu & \textsc{vi}  & 1060.890 &  too strong in the model  \\
   &              & 1082.265 &  too weak in model  \\
   &              & 1094.708 &  uncertain  \\
   &              & 1098.480 &  uncertain  \\
   &              & 1104.996 &  blend \Ion{Ga}{5}  \\
   &              & 1115.821 &  uncertain  \\
   &              & 1157.930 &  uncertain  \\
\hline
\end{longtable}
\noindent

\begin{longtable}{rlp{5.5cm}l}
\caption{\label{tab:Znlines}Identified Zn lines with $W_\lambda \ge 5$\,m{\AA} in model spectrum of \bd.} \\
\hline\hline
\noalign{\smallskip}
\multicolumn{2}{c}{Ion} & Wavelength\,/\,{\AA} & Comment \\ 
\noalign{\smallskip}
\hline
\endfirsthead
\caption{continued.}\\
\hline\hline
\noalign{\smallskip}
\multicolumn{2}{c}{Ion} & Wavelength\,/\,{\AA} & Comment \\
\noalign{\smallskip}
\hline
\noalign{\smallskip}
\endhead
\hline
\noalign{\smallskip}
\endfoot
\noalign{\smallskip}
Zn & \textsc{iv}   & 1239.119 & \\
   &               & 1281.296 & \\
Zn & \textsc{v}    & 1017.935 & blend ISM\\
   &               & 1023.521 & blend ISM\\
   &               & 1043.353 & \\
   &               & 1052.441 & blend ISM\\
   &               & 1053.278 & blend ISM\\
   &               & 1055.878 & blend ISM\\
   &               & 1056.330 & uncertain\\
   &               & 1058.185 & blend \Ion{Ga}{5}\\
   &               & 1061.472 & uncertain\\
   &               & 1061.656 & blend ISM\\
   &               & 1063.209 & blend ISM\\
   &               & 1063.979 & \\
   &               & 1066.547 & uncertain\\
   &               & 1068.284 & \\
   &               & 1069.674 & uncertain\\
   &               & 1069.764 & uncertain\\
   &               & 1071.501 & blend ISM\\
   &               & 1072.992 & blend ISM\\
   &               & 1074.241 & uncertain\\
   &               & 1075.171 & too strong in model\\
   &               & 1076.878 & \\
   &               & 1085.290 & uncertain\\
   &               & 1086.033 & blend \Ion{Ge}{5}\\
   &               & 1086.739 & blend ISM\\
   &               & 1088.709 & \\
   &               & 1090.831 & \\
   &               & 1094.088 & blend ISM\\
   &               & 1095.797 & uncertain\\
   &               & 1095.961 & uncertain\\
   &               & 1098.108 & \\
   &               & 1102.490 & uncertain\\
   &               & 1103.598 & too weak in model\\
   &               & 1104.199 & blend ISM \\
   &               & 1106.788 & \\
   &               & 1107.318 & \\
   &               & 1109.078 & \\
   &               & 1109.166 & \\
   &               & 1111.530 & \\
   &               & 1111.603 & \\
   &               & 1112.829 & \\
   &               & 1114.482 & \\
   &               & 1115.266 & \\
   &               & 1115.680 & \\
   &               & 1116.630 & too strong in model\\
   &               & 1116.842 & \\
   &               & 1117.466 & too weak in the model\\
   &               & 1118.778 & \\
   &               & 1119.950 & \\
   &               & 1120.101 & uncertain\\
   &               & 1120.325 & \\
   &               & 1121.109 & blend \Ion{Cr}{5}, \Ion{Fe}{6}\\
   &               & 1121.524 & uncertain\\
   &               & 1122.502 & blend ISM\\
   &               & 1123.127 & blend \Ion{Ga}{5}\\
   &               & 1124.718 & \\
   &               & 1125.019 1125.048 & \\
   &               & 1126.660 & \\
   &               & 1127.242 & \\
   &               & 1128.098 & blend \Ion{Ga}{5}\\
   &               & 1128.244 & uncertain\\
   &               & 1128.813 & \\
   &               & 1129.898 & \\
   &               & 1130.051 & \\
   &               & 1130.242 & \\
   &               & 1131.242 & \\
   &               & 1131.788 & \\
   &               & 1132.271 & \\
   &               & 1132.659 & blend \Ion{Co}{6}\\
   &               & 1133.031 & \\
   &               & 1133.128 & \\
   &               & 1133.278 & \\
   &               & 1133.498 & \\
   &               & 1135.324 & \\
   &               & 1135.588 & \\
   &               & 1136.311 & \\
   &               & 1136.603 & \\
   &               & 1136.986 & \\
   &               & 1137.625 & \\
   &               & 1138.248 & blend \Ion{Co}{6}\\
   &               & 1138.497 & \\
   &               & 1138.586 & \\
   &               & 1139.278 & uncertain\\
   &               & 1139.997 & uncertain\\
   &               & 1140.703 & \\
   &               & 1141.003 & \\
   &               & 1141.095 & \\
   &               & 1141.344 & uncertain\\
   &               & 1142.792 & \\
   &               & 1142.925 & \\
   &               & 1143.196 & \\
   &               & 1143.403 & blend \Ion{Ga}{5}\\
   &               & 1144.136 & \\
   &               & 1145.151 & \\
   &               & 1146.057 & \\
   &               & 1146.149 & \\
   &               & 1147.020 & \\
   &               & 1147.371 & \\
   &               & 1147.648 & \\
   &               & 1148.922 & \\
   &               & 1149.398 & \\
   &               & 1149.486 & \\
   &               & 1149.608 & \\
   &               & 1149.873 & blend \Ion{Ni}{6} \\
   &               & 1150.743 & \\
   &               & 1151.368 & \\
   &               & 1151.787 & uncertain\\
   &               & 1152.985 & \\
   &               & 1153.160 & \\
   &               & 1155.027 & blend \Ion{Fe}{7}\\
   &               & 1155.725 & \\
   &               & 1156.394 & \\
   &               & 1157.725 & uncertain\\
   &               & 1158.475 & blend \Ion{Ga}{5}\\
   &               & 1158.759 & \\
   &               & 1160.221 & \\
   &               & 1160.827 & uncertain\\
   &               & 1161.971 & \\
   &               & 1162.281 & \\
   &               & 1162.401 & uncertain\\
   &               & 1163.779 & uncertain\\
   &               & 1164.090 & \\
   &               & 1164.632 & \\
   &               & 1165.189 & blend \Ion{Ge}{5}\\
   &               & 1165.706 & \\
   &               & 1165.880 & too strong in model\\
   &               & 1168.302 & uncertain\\
   &               & 1169.200 & uncertain\\
   &               & 1169.301 & uncertain\\
   &               & 1170.105 & \\
   &               & 1170.885 & uncertain\\
   &               & 1171.106 & uncertain\\
   &               & 1171.422 & \\
   &               & 1171.801 & \\
   &               & 1171.951 & \\
   &               & 1172.038 & \\
   &               & 1173.366 & \\
   &               & 1173.823 & uncertain\\
   &               & 1173.892 & too strong in model\\
   &               & 1174.346 & \\
   &               & 1174.945 & uncertain\\
   &               & 1176.122 & \\
   &               & 1176.527 & too weak in model\\
   &               & 1176.868 1176.911  & \\
   &               & 1176.980 1177.016 1177.036 1177.087 & \\
   &               & 1178.639 & \\
   &               & 1178.759 & \\
   &               & 1179.145 & \\
   &               & 1179.969 1180.018 & \\
   &               & 1182.019 & uncertain\\
   &               & 1182.567 & \\
   &               & 1183.041 1183.158 & \\
   &               & 1183.314 & uncertain \\
   &               & 1185.619 1185.645 1185.676 & \\
   &               & 1185.898 1185.948 1185.961 & \\
   &               & 1186.057 & \\
   &               & 1186.447 & \\
   &               & 1187.706 & \\
   &               & 1189.072 & \\
   &               & 1189.331 & \\
   &               & 1190.003 & \\
   &               & 1190.376 & \\
   &               & 1192.014 & \\
   &               & 1192.703 1192.755 & \\  
   &               & 1193.846 & \\
   &               & 1195.745 & \\
   &               & 1198.795 & \\
   &               & 1200.639 & \\
   &               & 1201.961 & \\
   &               & 1202.128 & too strong in model \\
   &               & 1202.906 & \\
   &               & 1204.391 & \\
   &               & 1204.722 & \\
   &               & 1205.380 & too strong in the model \\
   &               & 1224.788 & \\
   &               & 1230.267 & \\
   &               & 1238.430 & \\
   &               & 1239.108 & \\
   &               & 1247.065 & \\
   &               & 1262.252 & \\
   &               & 1268.158 & \\
   &               & 1274.197 & \\
   &               & 1281.310 & \\
   &               & 1295.850 & too strong in model\\
   &               & 1302.786 & \\
   &               & 1318.204 & \\
   &               & 1344.241 & too strong in model\\
\hline
\end{longtable}
\noindent

\begin{longtable}{rlp{5.5cm}l}
\caption{\label{tab:Galines} Like \ta{tab:Znlines}, for Ga.} \\
\hline\hline
\noalign{\smallskip}
\multicolumn{2}{c}{Ion} & Wavelength\,/\,{\AA} & Comment \\ 
\noalign{\smallskip}
\hline
\endfirsthead
\caption{continued.}\\
\hline\hline
\noalign{\smallskip}
\multicolumn{2}{c}{Ion} & Wavelength\,/\,{\AA} & Comment \\
\noalign{\smallskip}
\hline
\noalign{\smallskip}
\endhead
\hline
\noalign{\smallskip}
\endfoot
\noalign{\smallskip}
Ga & \textsc{iv}  &    965.237 965.272  &  blend ISM\\
   &              &    981.831    &  blend ISM \\
   &              &   1003.780   &  blend ISM \\
   &              &   1004.367   &  \\
   &              &   1005.270   &  \\
   &              &   1009.849   &  blend ISM \\
   &              &   1010.080   &  blend ISM \\
   &              &   1014.822   &  uncertain\\
   &              &   1074.966   &  uncertain\\
Ga & \textsc{v}   &    943.583   &   \\
   &              &    962.084   &  uncertain\\
   &              &    967.324 967.404  &  blend \Ion{Ge}{6}\\  
   &              &    979.614   &  \\ 
   &              &    980.988   &   \\ 
   &              &    982.395   &  blend ISM \\ 
   &              &    984.078   &  \\ 
   &              &    990.138   &  uncertain\\
   &              &    997.855   &  blend ISM  \\ 
   &              &   1002.617   &  blend ISM  \\ 
   &              &   1009.928   &  blend ISM  \\ 
   &              &   1014.456   &  blend ISM  \\ 
   &              &   1014.868   &  uncertain\\
   &              &   1015.610   &  too strong in model\\ 
   &              &   1019.711   &  uncertain \\ 
   &              &   1032.375   &  blend ISM \\ 
   &              &   1033.492 1033.549 1033.580 & \\ 
   &              &   1034.822   &   \\ 
   &              &   1037.334   &  blend ISM \\ 
   &              &   1038.778   &  blend ISM \\ 
   &              &   1040.204   &  blend ISM \\ 
   &              &   1045.850   &  blend ISM \\ 
   &              &   1047.504   &  \\ 
   &              &   1050.453   &  \\ 
   &              &   1053.930   &  blend ISM \\ 
   &              &   1054.430   &  uncertain \\ 
   &              &   1054.563   &  blend \Ion{Ge}{5} \\ 
   &              &   1058.123   &   \\ 
   &              &   1062.677   &  \\ 
   &              &   1063.807   &  blend ISM \\ 
   &              &   1065.371   &  uncertain \\ 
   &              &   1068.593 1068.616  & too strong in model  \\ 
   &              &   1069.484 1069.530  & too strong in model \\
   &              &   1069.587   &  too strong in model \\  
   &              &   1071.123 1071.168  & \\ 
   &              &   1073.791   &   \\ 
   &              &   1074.911   &   uncertain\\
   &              &   1078.225   &  blend ISM \\ 
   &              &   1078.795   &  \\ 
   &              &   1079.587 1079.599  & \\ 
   &              &   1079.879 1079.925  & \\ 
   &              &   1080.474   &  blend ISM \\ 
   &              &   1080.988   &  uncertain \\ 
   &              &   1087.358   &   \\ 
   &              &   1088.068   &   \\ 
   &              &   1091.703   &   \\ 
   &              &   1094.355   &  blend ISM \\ 
   &              &   1094.739   &   \\ 
   &              &   1095.110   &   \\ 
   &              &   1100.401   &  uncertain \\ 
   &              &   1101.613   &  uncertain \\ 
   &              &   1102.767 1102.803 & too strong in model \\ 
   &              &   1103.047   &  too strong in model \\ 
   &              &   1104.936   &   \\ 
   &              &   1105.253   &  \\ 
   &              &   1105.620   &   \\ 
   &              &   1107.763   &   \\ 
   &              &   1109.829   &  blend ISM \\ 
   &              &   1115.561   &  blend ISM \\ 
   &              &   1118.018   &  \\ 
   &              &   1118.318   &  \\ 
   &              &   1120.260   &  blend ISM \\ 
   &              &   1123.154   &   \\ 
   &              &   1123.646   &   \\ 
   &              &   1126.393   &   \\ 
   &              &   1127.332   &   \\ 
   &              &   1127.726   &  \\ 
   &              &   1127.752   &  \\ 
   &              &   1128.082   &  blend \Ion{Zn}{5}\\
   &              &   1128.554   &  \\ 
   &              &   1129.152   &   \\ 
   &              &   1129.956   &   \\ 
   &              &   1131.452   &   \\ 
   &              &   1132.054   &   \\ 
   &              &   1132.157   &   \\ 
   &              &   1133.247   &  blend \Ion{Zn}{5}\\
   &              &   1133.903   &   \\ 
   &              &   1136.067   &   \\ 
   &              &   1138.187   &  \\ 
   &              &   1143.367   &  \\ 
   &              &   1145.974   &   \\ 
   &              &   1148.409   &   too strong in model \\ 
   &              &   1150.113   &   \\ 
   &              &   1150.219   &   \\ 
   &              &   1154.708   &   uncertain \\ 
   &              &   1155.976   &   \\ 
   &              &   1156.511   &    \\ 
   &              &   1157.729   &   uncertain \\ 
   &              &   1158.534   &   \\ 
   &              &   1160.847   &   uncertain\\ 
   &              &   1161.994   &   uncertain\\ 
   &              &   1162.048   &   uncertain\\ 
   &              &   1178.967   &   blend \Ion{Ni}{5} \\ 
   &              &   1180.958   &   uncertain\\ 
   &              &   1183.110   &   \\ 
   &              &   1183.656   &   \\ 
   &              &   1189.329   &   blend \Ion{Zn}{5}\\ 
   &              &   1190.179   &   uncertain \\ 
   &              &   1191.029   &   \\ 
   &              &   1193.061   &   \\ 
   &              &   1197.633   &   \\ 
   &              &   1265.454   &   \\ 
   &              &   1276.911   &   blend \Ion{Fe}{6} \\ 
   &              &   1283.615   &   \\ 
   &              &   1311.389   &   \\ 
Ga & \textsc{vi}  &    915.720   &   no observation \\ 
   &              &    919.117   &   blend ISM\\
   &              &    929.964   &   \\
   &              &    935.522   &   blend ISM\\
   &              &    945.329   &   blend ISM\\
   &              &    948.171   &   blend ISM\\
   &              &    953.738   &   blend ISM\\
   &              &    955.510   &  \\
   &              &    955.616   &  blend ISM \\
   &              &    956.648   &   blend ISM\\
   &              &    957.642   &   blend ISM\\
   &              &    960.172   &   blend ISM\\
   &              &    961.262   &    \\
   &              &    964.264 964.311 964.363 &  blend ISM \\
   &              &    964.569 964.647  &  blend ISM \\
   &              &    964.831 964.925  & blend ISM\\
   &              &    966.130   &   blend ISM\\
   &              &    966.255   &   blend ISM\\
   &              &    966.990   &   \\
   &              &    967.825   &   blend ISM\\
   &              &    968.107   &   \\
   &              &    970.064   &   \\
   &              &    974.853   &   \\
   &              &    975.165   &   \\
   &              &    975.342 975.396 & blend ISM \\
   &              &    976.133   &   blend ISM\\
   &              &    977.848   &   blend ISM\\
   &              &    978.897   &   uncertain\\
   &              &    979.298 979.383  & \\
   &              &    979.689   &   \\
   &              &    980.240   &   too strong in model\\
   &              &    980.489   &   uncertain\\
   &              &    980.580   &   blend ISM\\
   &              &    982.066   &   blend ISM\\
   &              &    983.110 983.160  & \\
   &              &    983.430 983.485  & uncertain\\
   &              &    983.630   &   \\
   &              &    984.009   &   blend ISM\\
   &              &    985.273   &   \\
   &              &    985.596   &   blend ISM\\
   &              &    985.812   &   blend ISM\\
   &              &    986.662   &   blend ISM\\
   &              &    987.862   &   blend ISM\\
   &              &    988.063   &   blend ISM\\
   &              &    989.169   &   blend ISM\\
   &              &    989.374   &   \\
   &              &    990.416   &   blend ISM\\
   &              &    990.639   &   \\
   &              &    992.053   &   blend ISM\\
   &              &    992.709   &   blend ISM\\
   &              &    993.094   &   \\
   &              &    993.640  993.654  & blend ISM \\
   &              &    994.051   &   blend ISM\\
   &              &    995.305   &   \\
   &              &    996.027   &   blend ISM\\
   &              &    996.309 996.391  & uncertain\\
   &              &    996.556   &   \\
   &              &    997.605   &   blend ISM\\
   &              &    999.083   &   blend ISM\\
   &              &    999.673   &   \\
   &              &    999.945   &   \\
   &              &   1000.117   &   \\
   &              &   1000.531   &   too strong in model\\
   &              &   1001.483   &   \\
   &              &   1001.821   &   blend ISM\\
   &              &   1002.376   &   blend ISM\\
   &              &   1002.985   &   uncertain\\
   &              &   1003.127 1003.147  & uncertain \\
   &              &   1003.390 1003.427  & uncertain \\
   &              &   1003.691   &   uncertain\\
   &              &   1004.170   &   blend ISM\\
   &              &   1004.355   &   \\
   &              &   1005.228   &   uncertain\\
   &              &   1006.156   &   uncertain\\
   &              &   1006.396   &   blend ISM\\
   &              &   1006.894 1006.951  & too strong in model \\
   &              &   1007.264   &   blend \Ion{Fe}{5}\\
   &              &   1007.511   &   blend \Ion{Fe}{5}\\
   &              &   1008.086   &   blend ISM\\
   &              &   1008.757   &   blend ISM\\
   &              &   1008.924 1009.049  &blend ISM\\
   &              &   1009.262   &   \\
   &              &   1009.512   &   blend ISM\\
   &              &   1009.743 1009.796 1009.806 & blend ISM \\
   &              &   1010.102   &   blend ISM\\
   &              &   1011.047   &   blend ISM\\
   &              &   1011.696  1011.698  & blend ISM  \\
   &              &   1012.260   &   blend ISM\\
   &              &   1013.620   &   blend ISM\\
   &              &   1014.434   &   blend ISM\\
   &              &   1015.598   &   too strong in model\\
   &              &   1015.782   &   \\
   &              &   1016.307   &   uncertain\\
   &              &   1017.003   &   \\
   &              &   1017.074   &   \\
   &              &   1018.785   &   \\
   &              &   1019.083   &   blend ISM\\
   &              &   1019.206   &   \\
   &              &   1020.741   &   blend ISM\\
   &              &   1022.008 1022.098  & uncertain  \\
   &              &   1027.809   &   uncertain \\
   &              &   1029.168   &   uncertain\\
   &              &   1030.457 1030.469  & uncertain  \\
   &              &   1037.061   &   blend ISM\\
   &              &   1037.419   &   blend ISM\\
   &              &   1038.800   &   blend ISM\\
   &              &   1042.324   &   blend \Ion{Fe}{5} \\
   &              &   1047.696   &   blend ISM\\
   &              &   1051.311   &   blend ISM\\
   &              &   1051.589   &   blend ISM\\
   &              &   1058.653   &   \\
   &              &   1058.931   &   \\
   &              &   1059.854 1058.931  &  \\
   &              &   1060.387   &   uncertain\\
   &              &   1061.790   &   blend ISM\\
   &              &   1063.972   &   uncertain\\
   &              &   1066.724   &   blend ISM\\
   &              &   1071.544   &   blend ISM\\
   &              &   1073.814   &   \\
   &              &   1076.760   &   \\
   &              &   1077.944   &   blend ISM\\
   &              &   1078.331   &   blend ISM\\
   &              &   1089.450   &   \\
   &              &   1099.109   &   \\
   &              &   1101.237   &   \\
   &              &   1105.669   &   \\
   &              &   1106.251   &   \\
   &              &   1121.262   &   uncertain\\
   &              &   1129.110   &   \\
   &              &   1138.016   &   \\
   &              &   1149.078   &   \\
   &              &   1237.472   &   \\
   &              &   1259.673   &   blend \Ion{Co}{6} \\ 
   &              &   1288.359   &   \\
\hline
\end{longtable}
\noindent

\begin{longtable}{rlp{5.5cm}l}
\caption{\label{tab:Gelines}Like \ta{tab:Znlines}, for Ge.} \\
\hline\hline
\noalign{\smallskip}
\multicolumn{2}{c}{Ion} & Wavelength\,/\,{\AA} & Comment \\ 
\noalign{\smallskip}
\hline
\endfirsthead
\caption{continued.}\\
\hline\hline
\noalign{\smallskip}
\multicolumn{2}{c}{Ion} & Wavelength\,/\,{\AA} & Comment \\
\noalign{\smallskip}
\hline
\noalign{\smallskip}
\endhead
\hline
\noalign{\smallskip}
\endfoot
\noalign{\smallskip}
Ge & \textsc{iv}  & 936.765   & blend ISM \\
   &              & 1189.028 & \\
   &              & 1229.839 & \\
   &              & 1494.889 & uncertain\\
Ge & \textsc{v}   & 942.717  &  blend ISM\\
   &              &  958.508 & \\
   &              &  965.501 & \\
   &              &  971.357 & blend ISM\\
   &              &  977.455 & \\
   &              &  984.923 & blend ISM\\
   &              &  986.767 & blend ISM\\
   &              &  988.132 & blend ISM\\
   &              &  990.668 & \\
   &              &  992.307 & blend ISM\\
   &              & 1004.380 & \\
   &              & 1004.938 & too strong in model\\
   &              & 1008.122 & blend ISM\\
   &              & 1016.667 & \\
   &              & 1033.107 & too strong in model\\
   &              & 1035.504 & uncertain\\
   &              & 1038.430 &  \\
   &              & 1042.127 & too strong in model\\
   &              & 1045.713 & \\
   &              & 1048.318 1048.371 1048.411 & uncertain\\
   &              & 1050.057 & blend ISM\\
   &              & 1054.590 & \\
   &              & 1058.932 & \\
   &              & 1068.430 & \\
   &              & 1069.132 & \\
   &              & 1069.420 & too strong in model\\
   &              & 1069.703 & \\
   &              & 1069.857 & uncertain\\
   &              & 1072.495 & \\
   &              & 1072.659 & \\
   &              & 1080.427 1080.484 1080.586 & blend ISM \\
   &              & 1086.653 & \\
   &              & 1087.855 & \\
   &              & 1089.491 1089.526 & \\
   &              & 1092.089 & \\
   &              & 1103.185 & uncertain\\
   &              & 1116.947 & \\
   &              & 1123.746 & too strong in model\\
   &              & 1125.424 & \\
   &              & 1139.187 & \\
   &              & 1163.400 & \\
   &              & 1165.259 & \\
   &              & 1176.690 & uncertain\\
   &              & 1222.300 & \\
Ge & \textsc{vi} & 911.098  &  no observation\\
   &              &  911.114 & no observation\\
   &              &  914.143 & no observation\\
   &              &  915.041 & no observation\\
   &              &  917.352 & blend ISM\\
   &              &  918.280 & blend ISM\\
   &              &  919.278 & blend ISM\\
   &              &  919.731 919.760 & blend ISM\\
   &              &  920.510 & \\
   &              &  921.084 & blend ISM\\
   &              &  921.780 & blend ISM\\
   &              &  923.486 & uncertain\\
   &              &  925.476 & \\
   &              &  926.822 & too strong in model\\
   &              &  928.136 &\\
   &              &  928.907 & \\
   &              &  929.428 & blend ISM\\
   &              &  929.631 & blend ISM\\
   &              &  930.081 & \\
   &              &  933.766 & blend ISM\\
   &              &  935.016 & \\
   &              &  935.912 & \\
   &              &  939.152 & blend ISM\\
   &              &  940.427 & \\
   &              &  942.474 & blend ISM\\
   &              &  942.567 & blend ISM\\
   &              &  944.851 & blend ISM\\
   &              &  946.589 & blend ISM\\
   &              &  947.937 & \\
   &              &  951.739 & \\
   &              &  952.415 & blend ISM\\
   &              &  953.132 & \\
   &              &  954.504 & blend ISM\\
   &              &  955.708 & blend ISM\\
   &              &  957.548 & blend ISM\\
   &              &  957.886 & \\
   &              &  958.100 & \\
   &              &  958.310 & uncertain\\
   &              &  960.102 & too strong in model\\
   &              &  964.813 & blend ISM\\
   &              &  965.203 & blend ISM\\
   &              &  965.914 & blend ISM\\
   &              &  967.300 & \\
   &              &  968.723 & \\
   &              &  969.010 & blend ISM\\
   &              &  969.171 & blend ISM\\
   &              &  969.568 &\\
   &              &  970.977 & blend ISM\\
   &              &  971.392 & blend ISM\\
   &              &  973.353 & blend ISM\\
   &              &  975.996 & \\
   &              &  979.258 & \\
   &              &  979.905 & blend ISM\\
   &              &  980.431 & blend ISM\\
   &              &  980.697 & too strong in model\\
   &              &  981.946 & blend ISM\\
   &              &  983.648 & \\
   &              &  988.180 & too strong in model\\
   &              &  990.049 & blend ISM\\
   &              &  990.848 & blend \Ion{Fe}{5} \\
   &              &  991.519 & blend ISM\\
   &              &  991.888 & blend ISM\\
   &              &  992.377 & blend \Ion{He}{2}\\
   &              &  993.015 & blend ISM\\
   &              &  995.289 & \\
   &              &  996.771 & too strong in model\\
   &              & 1002.276 & blend ISM\\
   &              & 1004.661 & \\
   &              & 1008.410 & blend ISM\\
   &              & 1011.518 & \\
   &              & 1013.528 & blend ISM\\
   &              & 1014.766 & \\
   &              & 1015.582 & too strong in model\\
   &              & 1016.817 & blend ISM\\
   &              & 1023.527 & blend ISM\\
   &              & 1031.263 1031.324 & blend ISM\\
   &              & 1033.152 & too strong in model\\
   &              & 1034.251 & uncertain\\
   &              & 1039.483 & too strong in model\\
   &              & 1039.890 & too strong in model\\
   &              & 1047.139 & too strong in model\\
   &              & 1047.207 & too strong in model\\
   &              & 1051.515 & blend ISM\\
   &              & 1053.196 & too strong in model\\
   &              & 1061.890 & \\
   &              & 1062.394 & \\
   &              & 1064.323 & \\
   &              & 1080.148 & too strong in model\\
   &              & 1080.510 & blend ISM\\
   &              & 1088.522 & \\
   &              & 1102.998 & too strong in model\\
   &              & 1103.103 & too strong in model\\
   &              & 1106.146 & too strong in model\\
   &              & 1113.177 & \\
   &              & 1121.431 & uncertain\\
   &              & 1127.159 & \\
   &              & 1148.327 & too strong in model\\
   &              & 1227.850 & too strong in model\\
   &              & 1228.870 & too strong in model\\
   &              & 1237.128 & \\
   &              & 1237.892 & too strong in model\\
   &              & 1251.447 & too strong in model\\
   &              & 1255.318 & \\
   &              & 1257.218 & blend \Ion{Mo}{5}\\
   &              & 1292.751 & too strong in model\\
   &              & 1300.093 & too strong in model\\
   &              & 1305.406 & too strong in model\\
   &              & 1391.639 & too strong in model\\
   &              & 1500.600 & \\
\hline
\end{longtable}
\noindent

\begin{longtable}{rlp{5.5cm}l}
\caption{\label{tab:Selines}Like \ta{tab:Znlines}, for Se.} \\
\hline\hline
\noalign{\smallskip}
\multicolumn{2}{c}{Ion} & Wavelength\,/\,{\AA} & Comment \\ 
\noalign{\smallskip}
\hline
\endfirsthead
\caption{continued.}\\
\hline\hline
\noalign{\smallskip}
\multicolumn{2}{c}{Ion} & Wavelength\,/\,{\AA} & Comment \\
\noalign{\smallskip}
\hline
\noalign{\smallskip}
\endhead
\hline
\noalign{\smallskip}
\endfoot
\noalign{\smallskip}
Se & \textsc{v}  &  943.957 & blend ISM  \\
   &             &  964.515 & blend ISM  \\
   &             & 1030.609 & \\
   &             & 1047.219 & too strong in model\\
   &             & 1059.951 & uncertain\\
   &             & 1094.691 & \\
   &             & 1151.016 &  \\
   &             & 1184.343 & \\
   &             & 1227.540 &  \\
   &             & 1249.048 & uncertain \\
   &             & 1264.063 & blend \Ion{Mn}{6} \\
   &             & 1426.676 & \\
   &             & 1433.568 & \\
   &             & 1445.567 & \\
   &             & 1447.283 & \\
   &             & 1447.408 & \\
   &             & 1451.653 & \\
   &             & 1454.292 & \\
   &             & 1473.253 & \\
   &             & 1730.014 & no observation\\      
   &             & 1736.835 & no observation\\      
   &             & 1740.038 & no observation\\
 \hline
\end{longtable}
\noindent

\begin{longtable}{rlp{5.5cm}l}
\caption{\label{tab:Brlines}Like \ta{tab:Znlines}, for Br.} \\
\hline\hline
\noalign{\smallskip}
\multicolumn{2}{c}{Ion} & Wavelength\,/\,{\AA} & Comment \\ 
\noalign{\smallskip}
\hline
\endfirsthead
\caption{continued.}\\
\hline\hline
\noalign{\smallskip}
\multicolumn{2}{c}{Ion} & Wavelength\,/\,{\AA} & Comment \\
\noalign{\smallskip}
\hline
\noalign{\smallskip}
\endhead
\hline
\noalign{\smallskip}
\endfoot
\noalign{\smallskip}
Br & \textsc{v}   &  945.815 & blend ISM\\
   &              & 1069.410 & \\
Br & \textsc{vi}  &  955.883 & blend ISM\\
   &              &  966.753 & blend ISM\\
   &              &  969.735 & \\
   &              &  981.423 & blend ISM\\
   &              & 1050.548 & \\
   &              & 1069.382 1069.432 &  \\
   &              & 1073.708 & \\
   &              & 1074.992 & \\
   &              & 1230.318 & \\
   &              & 1399.153 & uncertain\\
\hline
\end{longtable}
\noindent

\begin{longtable}{rlp{5.5cm}l}
\caption{\label{tab:Krlines}Like \ta{tab:Znlines}, for Kr.} \\
\hline\hline
\noalign{\smallskip}
\multicolumn{2}{c}{Ion} & Wavelength\,/\,{\AA} & Comment \\ 
\noalign{\smallskip}
\hline
\endfirsthead
\caption{continued.}\\
\hline\hline
\noalign{\smallskip}
\multicolumn{2}{c}{Ion} & Wavelength\,/\,{\AA} & Comment \\
\noalign{\smallskip}
\hline
\noalign{\smallskip}
\endhead
\hline
\noalign{\smallskip}
\endfoot
\noalign{\smallskip}
Kr & \textsc{vi} &  927.334  & blend ISM \\
   &              &  944.046 & uncertain\\
   &              &  965.093 & blend ISM\\
   &              &  980.411 & \\
   &              & 1002.748 & \\
   &              & 1045.238 & \\
   &              & 1061.064 & \\
Kr & \textsc{vii} & 918.444 & blend ISM  \\
   &              & 960.645 & blend ISM\\
   &              & 1197.166 & uncertain\\
\hline
\end{longtable}
\noindent

\begin{longtable}{rlp{5.5cm}l}
\caption{\label{tab:Srlines}Like \ta{tab:Znlines}, for Sr.} \\
\hline\hline
\noalign{\smallskip}
\multicolumn{2}{c}{Ion} & Wavelength\,/\,{\AA} & Comment \\ 
\noalign{\smallskip}
\hline
\endfirsthead
\caption{continued.}\\
\hline\hline
\noalign{\smallskip}
\multicolumn{2}{c}{Ion} & Wavelength\,/\,{\AA} & Comment \\
\noalign{\smallskip}
\hline
\noalign{\smallskip}
\endhead
\hline
\noalign{\smallskip}
\endfoot
\noalign{\smallskip}
Sr & \textsc{v} & 917.802 & blend ISM \\
   &              &  927.356 & blend ISM  \\
   &              &  928.353 & blend ISM  \\
   &              &  935.509 & blend ISM  \\
   &              &  936.808 & blend ISM  \\
   &              &  942.943 & blend ISM  \\
   &              &  946.530 & blend ISM  \\
   &              &  951.044 & blend ISM  \\
   &              &  951.159 & blend ISM  \\
   &              &  955.369 & blend \Ion{N}{4} \\
   &              &  957.714 & blend ISM\\
   &              &  962.378 & \\
   &              &  969.103 & blend ISM \\
   &              &  979.150 & \\
   &              &  985.408 & blend ISM \\
   &              & 1007.201 & uncertain\\
   &              & 1011.422 & \\
   &              & 1013.714 & blend ISM \\
   &              & 1020.002 & uncertain \\
   &              & 1020.439 & \\
   &              & 1030.445 & too strong in model\\
   &              & 1031.343 & blend ISM\\
   &              & 1038.990 & uncertain \\
   &              & 1041.940 & \\
   &              & 1056.104 & uncertain\\
   &              & 1065.215 & uncertain \\
   &              & 1070.578 & uncertain\\
   &              & 1114.594 & \\
   &              & 1114.876 & uncertain \\
   &              & 1132.353 & uncertain\\
   &              & 1141.221 & uncertain\\
   &              & 1152.104 & uncertain\\
   &              & 1154.871 & \\
   &              & 1163.040 & \\
   &              & 1164.173 & blend \Ion{Zn}{5} \\
   &              & 1175.115 & uncertain\\
   &              & 1200.728 & blend ISM\\
   &              & 1238.652 & blend \Ion{Ni}{5} \\
   &              & 1280.995 & \\
   &              & 1281.911 & \\
   &              & 1311.334 & \\
   &              & 1311.781 & blend \Ion{Fe}{5} \\
   &              & 1387.288 & \\
   &              & 1415.404 & \\
   &              & 1447.665 & \\
   &              & 1459.369 & \\
   &              & 1472.784 & \\
Sr & \textsc{vi}  & 912.3760 & no observation \\
\hline
\end{longtable}
\noindent

\begin{longtable}{rlp{5.5cm}l}
\caption{\label{tab:Zrlines}Like \ta{tab:Znlines}, for Zr.} \\
\hline\hline
\noalign{\smallskip}
\multicolumn{2}{c}{Ion} & Wavelength\,/\,{\AA} & Comment \\ 
\noalign{\smallskip}
\hline
\endfirsthead
\caption{continued.}\\
\hline\hline
\noalign{\smallskip}
\multicolumn{2}{c}{Ion} & Wavelength\,/\,{\AA} & Comment \\
\noalign{\smallskip}
\hline
\noalign{\smallskip}
\endhead
\hline
\noalign{\smallskip}
\endfoot
\noalign{\smallskip}
Zr & \textsc{v}   & 1200.802  & blend ISM\\
   &              & 1332.065  & \\
Zr & \textsc{vi}  & 955.500  &  \\
   &              & 1040.904  & \\
   &              & 1040.995  & \\
   &              & 1044.483  & uncertain\\
   &              & 1050.580  & blend \Ion{Br}{6}\\
   &              & 1053.548  & \\
   &              & 1064.818  & blend ISM\\
   &              & 1068.836  & \\
   &              & 1072.877  & \\
   &              & 1073.197  & uncertain\\
   &              & 1074.554  & \\
   &              & 1081.130  & \\
   &              & 1081.384  & blend ISM\\
   &              & 1088.439  & uncertain\\
   &              & 1095.491  & \\
   &              & 1099.591  & \\
   &              & 1101.742  & \\
   &              & 1108.491  & blend ISM\\
   &              & 1113.736  & uncertain\\
   &              & 1114.481  & \\
   &              & 1118.689  & too strong in model\\
   &              & 1129.371  & \\
   &              & 1134.606  & uncertain\\
   &              & 1142.550  & \\
   &              & 1143.933  & \\
   &              & 1150.774  & \\
   &              & 1151.571  & \\
   &              & 1158.582  & uncertain\\
   &              & 1161.639  & \\
   &              & 1314.034  & \\
   &              & 1417.865  & \\
   &              & 1514.568  & \\
   &              & 1521.699  & \\
   &              & 1529.396  & \\
   &              & 1536.035  & uncertain\\
   &              & 1538.423  & \\
   &              & 1541.255  & uncertain\\
   &              & 1591.799  & \\
   &              & 1604.549  & uncertain\\
   &              & 1645.326  & \\
   &              & 1663.952  & \\
   &              & 1679.018  & uncertain\\
   &              & 1682.241  & \\
   &              & 1683.302  & uncertain\\
   &              & 1733.091  & no observation\\
   &              & 1733.937  & no observation\\
   &              & 1741.948  & no observation\\
   &              & 1749.350  & no observation\\
Zr & \textsc{vii} & 1233.578  & \\
   &              & 1234.964  & \\
   &              & 1376.633  & \\
   &              & 1469.098  & uncertain\\
\hline
\end{longtable}
\noindent
\

\begin{longtable}{rlp{5.5cm}l}
\caption{\label{tab:Molines}Like \ta{tab:Znlines}, for Mo.} \\
\hline\hline
\noalign{\smallskip}
\multicolumn{2}{c}{Ion} & Wavelength\,/\,{\AA} & Comment \\ 
\noalign{\smallskip}
\hline
\endfirsthead
\caption{continued.}\\
\hline\hline
\noalign{\smallskip}
\multicolumn{2}{c}{Ion} & Wavelength\,/\,{\AA} & Comment \\
\noalign{\smallskip}
\hline
\noalign{\smallskip}
\endhead
\hline
\noalign{\smallskip}
\endfoot
\noalign{\smallskip}
Mo & \textsc{vi}  & 995.806  & uncertain\\
   &              & 1038.640 & blend ISM\\
   &              & 1047.182 & uncertain \\
   &              & 1479.168 & \\
   &              & 1595.435 & \\
\hline
\end{longtable}
\noindent

\begin{longtable}{rlp{5.5cm}l}
\caption{\label{tab:Inlines}Like \ta{tab:Znlines}, for In.} \\
\hline\hline
\noalign{\smallskip}
\multicolumn{2}{c}{Ion} & Wavelength\,/\,{\AA} & Comment \\ 
\noalign{\smallskip}
\hline
\endfirsthead
\caption{continued.}\\
\hline\hline
\noalign{\smallskip}
\multicolumn{2}{c}{Ion} & Wavelength\,/\,{\AA} & Comment \\
\noalign{\smallskip}
\hline
\noalign{\smallskip}
\endhead
\hline
\noalign{\smallskip}
\endfoot
\noalign{\smallskip}
In & \textsc{v}   &  933.577  & blend ISM\\
   &              &  940.079  & uncertain \\
   &              &  942.218  & uncertain \\
   &              & 1101.860  & uncertain\\
   &              & 1122.517  & blend ISM\\
   &              & 1135.588  & uncertain\\
   &              & 1136.347  & \\
   &              & 1137.787  & \\
   &              & 1148.852  & \\
   &              & 1151.723  & \\
   &              & 1153.836  & \\
   &              & 1156.652  & \\
   &              & 1160.561  & \\
   &              & 1168.056  & \\
   &              & 1177.447  & \\
   &              & 1181.329  & \\
   &              & 1183.049  & \\
   &              & 1190.489  & blend ISM\\
   &              & 1191.583  & \\
   &              & 1192.278  & \\
   &              & 1192.541  & \\
   &              & 1196.281  & \\
   &              & 1199.170  & \\
   &              & 1200.843  & blend ISM\\
   &              & 1210.126  & uncertain\\
   &              & 1228.000  & uncertain\\
   &              & 1228.483  & uncertain\\
   &              & 1238.448  & \\
   &              & 1241.025  & \\
   &              & 1241.299  & \\
   &              & 1242.210  & \\
   &              & 1243.632  & blend \Ion{Ni}{5} \\
   &              & 1252.836  & blend \Ion{Fe}{6}\\
   &              & 1256.570  & uncertain \\
   &              & 1276.318  & uncertain\\
   &              & 1278.758  &  \\
   &              & 1285.468  & uncertain\\
   &              & 1289.800  & \\
   &              & 1290.449  & \\
   &              & 1292.930  & \\
   &              & 1295.035  & uncertain\\
   &              & 1296.427  & \\
   &              & 1315.139  & \\
   &              & 1317.671  & uncertain\\
   &              & 1320.468  & uncertain\\
   &              & 1334.123  & \\
   &              & 1339.599  & \\
   &              & 1355.458  & uncertain\\
\hline
\end{longtable}
\noindent

\begin{longtable}{rlp{5.5cm}l}
\caption{\label{tab:Telines}Like \ta{tab:Znlines}, for Te.} \\
\hline\hline
\noalign{\smallskip}
\multicolumn{2}{c}{Ion} & Wavelength\,/\,{\AA} & Comment \\ 
\noalign{\smallskip}
\hline
\endfirsthead
\caption{continued.}\\
\hline\hline
\noalign{\smallskip}
\multicolumn{2}{c}{Ion} & Wavelength\,/\,{\AA} & Comment \\
\noalign{\smallskip}
\hline
\noalign{\smallskip}
\endhead
\hline
\noalign{\smallskip}
\endfoot
\noalign{\smallskip}
Te & \textsc{vi}  &  951.021 & blend ISM\\
   &              & 1071.414 & \\
   &              & 1242.023 & uncertain\\
   &              & 1267.986 & \\
   &              & 1313.874 & \\
\hline
\end{longtable}
\noindent

\begin{longtable}{rlp{5.5cm}l}
\caption{\label{tab:Ilines}Like \ta{tab:Znlines}, for I.} \\
\hline\hline
\noalign{\smallskip}
\multicolumn{2}{c}{Ion} & Wavelength\,/\,{\AA} & Comment \\ 
\noalign{\smallskip}
\hline
\endfirsthead
\caption{continued.}\\
\hline\hline
\noalign{\smallskip}
\multicolumn{2}{c}{Ion} & Wavelength\,/\,{\AA} & Comment \\
\noalign{\smallskip}
\hline
\noalign{\smallskip}
\endhead
\hline
\noalign{\smallskip}
\endfoot
\noalign{\smallskip}
I & \textsc{vi} &  911.192  & no observation\\
  &             &  919.210  & blend ISM\\
  &             &  970.448  & uncertain\\
  &             &  987.381  & blend ISM\\
  &             &  989.005  & blend ISM\\
  &             & 1000.999  & uncertain\\
  &             & 1045.423  & \\
  &             & 1053.389  & blend ISM\\
  &             & 1057.530  & \\
  &             & 1120.301  & blend ISM\\
  &             & 1121.218  & uncertain\\
  &             & 1137.370  & uncertain\\
  &             & 1153.262  & too strong in model\\
  &             & 1185.111  & uncertain \\
  &             & 1191.601  & \\
  &             & 1195.359  & uncertain\\
  &             & 1395.979  & \\
\hline
\end{longtable}
\noindent

\begin{longtable}{rlp{5.5cm}l}
\caption{\label{tab:Xelines}Like \ta{tab:Znlines}, for Xe.} \\
\hline\hline
\noalign{\smallskip}
\multicolumn{2}{c}{Ion} & Wavelength\,/\,{\AA} & Comment \\ 
\noalign{\smallskip}
\hline
\endfirsthead
\caption{continued.}\\
\hline\hline
\noalign{\smallskip}
\multicolumn{2}{c}{Ion} & Wavelength\,/\,{\AA} & Comment \\
\noalign{\smallskip}
\hline
\noalign{\smallskip}
\endhead
\hline
\noalign{\smallskip}
\endfoot
\noalign{\smallskip}
Xe & \textsc{vi}   & 1080.080 & blend \Ion{Ge}{6}\\
  &                & 1091.630 & uncertain\\
  &                & 1136.410 & uncertain \\
Xe & \textsc{vii}  & 912.875 & no observation\\
  &                &  920.861 & blend ISM\\
  &                &  942.152 & \\
  &                &  943.218 & \\
  &                &  970.177 & uncertain\\
  &                &  995.510 & \\
  &                &  997.407 & blend \Ion{Fe}{5} \\
  &                & 1071.226 & uncertain\\
  &                & 1077.110 & blend ISM\\
  &                & 1093.781 & blend ISM\\
  &                & 1243.565 & \\
  &                & 1460.856 & uncertain\\
\hline
\end{longtable}
\noindent

\begin{longtable}{rlp{5.5cm}l}
\caption{\label{tab:Balines}Like \ta{tab:Znlines}, for Ba.} \\
\hline\hline
\noalign{\smallskip}
\multicolumn{2}{c}{Ion} & Wavelength\,/\,{\AA} & Comment \\ 
\noalign{\smallskip}
\hline
\endfirsthead
\caption{continued.}\\
\hline\hline
\noalign{\smallskip}
\multicolumn{2}{c}{Ion} & Wavelength\,/\,{\AA} & Comment \\
\noalign{\smallskip}
\hline
\noalign{\smallskip}
\endhead
\hline
\noalign{\smallskip}
\endfoot
\noalign{\smallskip}
Ba & \textsc{vi}   & 937.595  & blend ISM \\
Ba & \textsc{vii}  & 924.898 & blend ISM \\
   &               &  943.102 & blend ISM\\
   &               &  993.411 & \\
   &               & 1074.937 & \\
   &               & 1255.520 & uncertain\\
   &               & 1465.045 & \\
Ba & \textsc{viii} & 921.761 & uncertain\\
   &               &  941.168 & uncertain\\
   &               &  952.762 & blend ISM\\
   &               &  961.679 & blend ISM\\
   &               & 1013.130 & blend ISM\\
   &               & 1039.555 & \\
   &               & 1048.339 & blend ISM\\
   &               & 1074.911 & \\
   &               & 1083.072 & \\
   &               & 1113.140 & uncertain\\
\hline
\end{longtable}
\noindent

\twocolumn

\bsp	
\label{lastpage}
\end{document}